\documentclass[12pt,a4paper]{article}
\usepackage{empheq}
\usepackage{amsmath,amscd}
\numberwithin{equation}{section}
\usepackage{mathrsfs}
\usepackage{empheq}
\usepackage{amsfonts}
\usepackage{graphicx}
\usepackage{amssymb,scalerel,stackengine}
\usepackage[utf8]{inputenc}
\usepackage{color}
\usepackage{parskip}
\usepackage[table]{xcolor}
\usepackage{float}
\usepackage{enumitem}
\setlist[description]{font=\normalfont\itshape}
\usepackage[most]{tcolorbox}
\newtcbox{\othermathbox}[1][]{nobeforeafter, math upper, tcbox raise base, 
	enhanced, rounded corners, colback=black!5, colframe=black}
\usepackage{authblk}   
\usepackage[margin=1in]{geometry}
\usepackage{graphicx}
\usepackage[leftcaption]{sidecap}
\usepackage[font=small,labelfont=bf]{caption}
\usepackage{subcaption}
\usepackage{here}
\usepackage[colorlinks=true,allcolors=blue]{hyperref}
\usepackage{cite}
\usepackage{ragged2e}
\usepackage{etoolbox}
\usepackage{tikz}
\usetikzlibrary{positioning,arrows.meta,calc}
\usepackage{float}
\apptocmd{\thebibliography}{\justifying}{}{}

\let\OLDthebibliography\thebibliography
\renewcommand\thebibliography[1]{
	\OLDthebibliography{#1}
	\setlength{\parskip}{0pt}
	\setlength{\itemsep}{4.85pt plus 0.3ex}
}

\usepackage{comment}
\def\cv{{\check{v}}}
\def\la{{\lambda}}
\def\th{{\theta}}
\def\Th{{\Theta}}

\def\hA{{\hat A}}

\def\he{{\hat e}}

\def\hk{{\hat \kappa}}

\def\hl{{\hat \ell}}

\def\hn{{\hat n}}
\def\hx{{\hat x}}

\def\pd{{\partial}}
\def\cd{{\nabla}}
\def\bg{{\bar{g}}}
\def\bl{{\bar{\ell}}}
\def\bV{{\overline{V}}}
\def\cC{{\Psi}}

\def\bD{{\overline{D}}}
\def\de{{\delta}}
\def\cF{{\mathcal F}}

\def\cH{{\mathcal H}}

\def\cL{{\mathcal L}}
\def\cS{{\mathcal S}}

\def\cY{{\mathcal Y}}

\def\bgamma{{\bar \gamma}}
\def\hell{{\hat \ell}}
\def\eps{{\epsilon}}

\def\mn{{\mu\nu}}
\def\deq{\mkern5mu\dot{=}\mkern5mu}

\def\bq{{\overline{q}}}

\def\cO{{\mathcal{O}}}
\def\bY{{\bar{Y}}}

\def\bt{{\bar{t}}}
\def\bsi{{\bar{\psi}}}
\def\bW{{\overline{\Omega}}}
\def\bk{{\overline{\kappa}}}

\def\hh{{\mathsf{h}}}

\def\hv{{\hat{v}}}
\def\hth{{\hat{\theta}}}
\def\vth{{\vartheta}}
\def\bw{{\overline{\omega}}}
\def\tlg{{\widetilde{g}}}

\def\ddhv{\frac{\dd}{\dd \hv}}
\def\af{{\footnotesize \text{aff}}}
\def\kerr{{\footnotesize\text{Kerr}}}
\def\dyn{{\footnotesize\text{dyn}}}

\def\Dv{{\mathcal{D}}}
\def\bDv{{\overline{\mathcal{D}}\,}}

\def\tlv{{\tilde{v}}}

\newcommand\po[1]{{\footnotesize {#1}}}
\newcommand\bo[1]{{\footnotesize {(#1)}}}

\newcommand\dd{\text{d}}

\def\ie{{\textit{i.e. }}}
\def\eg{{\textit{e.g. }}}

\newcommand\tl[1]{{\tilde #1}}
\newcommand\avg[1]{{\langle #1\rangle}}
\newcommand\osc[1]{{\langle #1\rangle}_\text{osc}}

\title{\bf Horizon flux-balance laws \\ in the multiscale perturbation}

\author[1,2]{Ali Seraj}

\affil[1]{\small\textit{School of Quantum Physics and Matter, Institute for Research in Fundamental Sciences\newline  (IPM),P.O.Box 19395-5531, Tehran, Iran}}
\affil[2]{\small \textit{Leuven Gravity Institute, KU Leuven, Celestijnenlaan 200D box 2415, 3001 Leuven, Belgium}\\
	\texttt{\href{ali\_seraj@ipm.ir}{ali\_seraj@ipm.ir}}}

\date{}
\begin{document}
	\maketitle
\begin{abstract}
Black-hole event horizons obey a set of evolution equations governing their intrinsic and extrinsic geometry. We study these equations in a \textit{two-timescale} perturbative expansion about a Kerr background and derive strong constraints on the coarse-grained horizon dynamics. At leading order, the coarse-grained linear shear vanishes, while the horizon exhibits an \textit{adiabatic rigidity}: the leading corrections to its angular velocity and inaffinity remain uniform on each horizon cut, while evolving on the slow timescale. We then provide a systematic procedure for transforming a perturbative bulk solution, given for example in Lorenz gauge, to an ingoing Newman--Unti gauge adapted to the horizon, allowing the perturbed horizon geometry to be extracted directly from the bulk metric. Finally, we formulate black-hole conservation laws associated with horizon symmetries, including energy, dynamical entropy, and angular momentum. We expand the charges through second order and their fluxes through third order in perturbation theory, providing a framework for future applications to horizon absorption and backreaction in extreme mass-ratio inspirals.
\end{abstract}

\tableofcontents

\section{Introduction}

The geometry and evolution of null hypersurfaces have a long history in
general relativity because of their importance in describing wavefronts of gravitational radiation, black hole event horizons, and boundaries of asymptotically flat spacetimes. In the early 1960s, Sachs introduced a set of optical scalars
characterizing null geodesic congruences and derived their propagation equations. They provide a geometric interpretation of certain spin coefficients and their associated equations in the
Newman--Penrose formalism, which was developed at approximately the same time~\cite{Sachs:1961zz,Newman:1961qr,Sachs:1962wk}.
Applied to black-hole event horizons, these equations underlie the laws of black-hole mechanics
\cite{Bardeen:1973gs} and the analysis by Hawking and Hartle of
energy and angular-momentum flow into a perturbed rotating black
hole~\cite{Hawking:1972hy}. Building on the geometric formulation of Hajicek, Damour
developed a hydrodynamic description of the horizon by showing that its
tangential evolution equation takes the form of a Navier--Stokes-type
equation
\cite{Hajicek1974can,Hajicek1975stationary,Damour:1978cg,Damour:1979wya}.
This interpretation of the horizon as an effective fluid plays a central role in the membrane paradigm developed by Price
and Thorne \cite{Price:1986yy,Thorne:1986iy}.
These developments provide a set of evolution equations governing
the intrinsic and extrinsic geometry of a null horizon, nicely
reviewed in~\cite{Gourgoulhon:2005ng}. 

More recently, covariant phase-space and null-boundary formulations have
related the gravitational constraints on null hypersurfaces to boundary
symmetries, charges, and flux-balance laws
\cite{Hopfmuller:2016scf,Hopfmuller:2018fni,Chandrasekaran:2018aop,
	Donnay:2019jiz,Chandrasekaran:2021hxc,Adami:2021nnf,
	Ciambelli:2023mir,Odak:2023pga,Chandrasekaran:2023vzb,Ashtekar:2021kqj,Ashtekar:2021wld,Ashtekar:2024stm}.
These approaches provide a natural framework for interpreting the
Raychaudhuri and Damour equations as gravitational conservation laws on a
null boundary and for identifying the associated charges and fluxes.

A complementary approach to black-hole horizon geometry is provided by the
isolated- and dynamical-horizon frameworks~\cite{Ashtekar:2004cn}.
Bondi-like coordinates and Newman--Penrose frames adapted to a neighborhood
of the horizon have been developed, in particular, in
\cite{Krishnan:2012bt}. Explicit constructions for Kerr and Kerr--Newman
are developed in
\cite{Fletcher:2003kpg,Scholtz:2017ttf,Kofron:2024taq,Flandera:2025jbn}. These coordinates play near the horizon a
role analogous to Bondi coordinates near null infinity: the intrinsic and extrinsic geometry of the
null hypersurface can be read off directly from the metric coefficients.

In extreme mass-ratio inspirals (EMRIs), the mass $m$ of the secondary
is much smaller than the mass $M$ of the primary black hole. Einstein's
equations can therefore be treated perturbatively in the small mass ratio
$\eps=m/M\ll1$. In the limit $\eps\to0$, the secondary follows a geodesic
of the background black-hole spacetime~\cite{Poisson:2011nh}. For small but finite $\eps$, however, its motion is governed by the
gravitational self-force, whose dissipative component drives a slow
evolution of the orbital parameters~\cite{Barack:2018yvs,Pound:2021qin}.
The problem is intrinsically multiscale, involving fast orbital timescales
and a much longer radiation-reaction timescale. A two-timescale expansion provides a natural framework for
describing this evolution~\cite{Hinderer:2008dm,Pound:2010pj}.

Radiation generated by the secondary is carried both to null infinity and
through the event horizon. Both contributions enter the leading adiabatic
balance laws governing the inspiral. 
Horizon absorption changes the mass and angular momentum of the primary,
and the resulting slow evolution of the primary's parameters contributes
at post-adiabatic order to the orbital dynamics of the secondary and the observed waveform at infinity~\cite{Poisson:2004cw,Miller:2020bft,Mathews:2025nyb}.
A systematic description of the intrinsic horizon geometry and of the
gravitational fluxes crossing it is therefore an important ingredient in
high-order EMRI perturbation theory, and waveform modeling for future gravitational-wave experiments such as LISA.

In this paper, we combine these different perspectives to study
flux-balance laws at the event horizon of the primary black hole in an EMRI.
We first analyze the horizon evolution equations in a two-timescale
perturbative expansion about Kerr and derive strong constraints on the
coarse-grained horizon dynamics. At leading order, the coarse-grained
linear shear vanishes, while the leading corrections to the horizon angular
velocity and inaffinity remain uniform on each horizon cut and evolve only
on the slow timescale, exhibiting a form of \textit{adiabatic rigidity}.
We then provide a systematic procedure for transforming a perturbative bulk
solution in a generic user gauge, such as Lorenz gauge, to an ingoing
Newman--Unti (INU) gauge adapted to the event horizon, from which the
perturbed horizon geometry can be extracted directly. Finally, using the
covariant phase-space formulation, we construct conservation laws associated
with horizon symmetries, including charges identified with black-hole's energy, dynamical entropy, and
angular momentum. We expand the corresponding charges through second order
and their fluxes through third order in the mass-ratio expansion, providing
the ingredients needed for future applications to horizon absorption and
backreaction in EMRIs. Figure~\ref{fig:summary} at the end of the paper provides a summary of the paper, and highlights the main results.

\paragraph{Notations and conventions.}
The following conventions will be used:
\begin{itemize}[leftmargin=*,topsep=0pt]
	\item We use natural units in which $G=c=1$.
	
	\item The symbol $\deq$ denotes \textit{equality on the horizon}; both
	sides of an equation containing $\deq$ are therefore evaluated on
	$\cH$. 
	
	\item Three manifolds are involved: the four-dimensional spacetime
	$\cal M$, the three-dimensional horizon $\cH$, and the two-dimensional
	cuts $\cS$ of the horizon. Greek indices $\mu,\nu,\ldots$ denote spacetime
	components, lowercase Latin indices $a,b,\ldots$ denote components
	intrinsic to $\cH$, and uppercase Latin indices $A,B,\ldots$ denote
	components intrinsic to its cuts $\cS$.
	
	\item 	An overbar denotes the value of a quantity on the background
	spacetime, while perturbative quantities are labeled by an appropriate
	subscript or superscript. For example, for a dynamical quantity $f$, we denote its background value by $\bar f$ and its $n$-th order perturbation by $f_\po{n}$ or equivalently $f^\bo{n}$. 
	
	\item In the exact horizon geometry, indices on quantities such as
	$\sigma_{AB}$ and $\omega_A$ are raised and lowered with the exact
	induced metric $q_{AB}$. In perturbation theory, indices on perturbative
	coefficients such as $\sigma^\bo{1}_{AB}$ and $\omega_\po{1}^{A}$ are
	raised and lowered with the background metric $\bq_{AB}$.
	\item The same symbol will be used to denote a vector and its dual one-form, \eg $\ell=\ell^\mu\pd_\mu$ and $\ell=\ell_\mu dx^\mu$. This will not produce any confusion.
\end{itemize}
\section{Ingoing Newman-Unti (INU) coordinates}\label{sec: INU}
In this section, we recall the construction of the ingoing Newman-Unti (INU) coordinate system $(v,s,x^A)$, suitable to describe the geometry of spacetime near the horizon of a black hole. A gauge-fixed version of this coordinate system was used by Carter to study stationary black holes and then generalized to dynamical black holes by Price and Thorne~\cite{carter2010republication,Price:1986yy}. In this work, we will mostly work with the INU coordinates, but we will also address how to gauge fix to Carter coordinates, and use the latter to describe charges and fluxes on the horizon. 

The horizon $\cH$ is a null hypersurface, and hence its normal $\ell^\mu$ is null $g_\mn \ell^\mu \ell^\nu=0$. The normal is simultaneously tangent to $\cH$ and generates null geodesics on $\cH$, so that $\ell^\nu \cd_\nu \ell^\mu =\kappa \ell^\mu$, where $\kappa$ is known as the \textit{inaffinity}. In fact, $\ell^\mu$ is defined up to an arbitrary scaling under which
\begin{align}\label{scaling}
	\hl^\mu=\alpha(x) \ell^\mu\,,\qquad  \hk= (\pd_\ell+ \kappa)\alpha\,,\qquad \alpha>0\,,
\end{align}  
where $\pd_\ell=\ell^\mu \pd_\mu=\ell^a\pd_a$. We can therefore pick a specific representative $\ell$ by fixing the value of the inaffinity. For a given $\hk(v,x^A)$, the first-order ODE $(\pd_\ell+ \kappa)\alpha=\hk$ always has a solution. We will come back to this point later.

Now consider a coordinate system $x^a= (v,x^A)$ on the horizon. $v$ is an advanced time parameter, used to foliate the horizon into two-dimensional cuts $\cS_v$ with spherical topology, while $x^A$ is a coordinate system on the cuts. Once a representative $\ell$ is chosen, we define $v$ so that 
\begin{align}\label{V def}
	\pd_\ell(v)\deq 1\quad \implies \quad	\ell\deq \pd_v+V^A\pd_A\,.
\end{align}
$V^A$ defines a velocity field on the horizon and is analogous to the shift vector in a 3+1 splitting of spacetime~\cite{Damour:1978cg,Damour:1979wya}. 
After imposing $\pd_\ell (v)\deq 1$, the choice of the scaling of the generator,  the coordinates $x^A$ on the cuts, as well as the location of the initial cut $S_0$ at $v=0$ constitute the gauge freedom on the horizon. These will be related to the symmetries of the phase space of geometries under consideration. While most of our discussion remains covariant in the $x^A$ coordinates, we may take advantage of the axial symmetry of the background Kerr black hole to define the usual spherical coordinates $(\vth,\varphi)$, such that  $\phi=\pd_\varphi$ is a Killing vector of the background metric.  In this case, $\bV=\bW \,\phi$, where $\bW$ is the angular velocity of the Kerr horizon.

A special choice for the horizon coordinates is the \textit{Carter coordinates} $(\hv,\hx^A)$.   We first rescale the generator $\ell\to\hat{\ell}$ such that the inaffinity is fixed to the surface gravity of the background, \ie $\hk=\bk$, and define $\hv$ by $\pd_\hl(\hv)=1$. We refer to this as \textit{Killing} parametrization, because on the background geometry $\hat\ell$ coincides with the asymptotically normalized Killing generator of the Kerr horizon. We then choose corotating angular coordinates satisfying
$\pd_\hl (\hx^A)=0$. Together, these conditions imply
\begin{align}
	\hat\ell=\pd_\hv.
\end{align}

To extend the coordinate system off the horizon, we consider an ingoing congruence of null geodesics, with tangent $n$, passing through the horizon. The congruence is assumed to be twist-free and a gradient, \ie there exists a function $f$ such that $n=\dd f$. The function $f$ is preserved by the flow of $n$ as $n\cdot \cd f=n\cdot n=0$. Therefore, if we identify $f=-v$ at the horizon, we find that $n=-\dd v$ everywhere. Moreover, let $s$ be an outgoing affine parameter so that $n=-\pd_s$, and identify $s=0$ with the location of the horizon. Horizon coordinates $(v,x^A)$ are extended outside the horizon such that they remain constant along integral curves of $n$. The congruence foliates the spacetime into $v=$const hypersurfaces crossing the horizon. Note that the choice of the horizon coordinates constrains the transversal vector $n$ through its orthogonality to the horizon basis $(\ell,e_A=\pd_A)$ given by $\ell\cdot n=-1$, $e_A\cdot n=0$. 

The coordinate system $(v,s,x^A)$ thus defined will be called the \textit{ingoing Newman-Unti} (INU) coordinates, in analogy with its counterpart construction at null infinity~\cite{Newman:1962cia}. The general metric in the INU coordinates satisfies the gauge conditions
\begin{align}\label{INU gauge}
	g_{ss}=g_{sA}=0\,,\qquad g_{vs}=1 \quad \Leftrightarrow\quad g^{vv}=g^{vA}=0\,,\quad g^{vs}=1\,,
\end{align}
and is written as
\begin{align}\label{INU metric}
			d{s}^2&= dv (-F dv +2ds)+{g}_{AB}(dx^A-U^A dv)(dx^B-U^B dv)\,.
\end{align}
In particular, the inverse metric takes the simple form (displaying only the upper corner)
\begin{align}
	g^\mn&	=
	\begin{pmatrix}
		0 & 1 & 0 \\[2mm]
		 & F & U^B \\[2mm]
		 & & g^{AB}
	\end{pmatrix}.
\end{align}
In terms of the INU coordinates, the null pair $\ell,n$ take the form
\begin{align}
	\ell&=\pd_v+\frac{F}{2}\pd_s +U^A\pd_A\,,& n&=-\pd_s\,,\\
	\ell&=ds-\frac{F}{2}dv& n&=-dv\,.
\end{align}

The variables $F,U^A,g_{AB}$ can be expanded near the horizon at $s=0$ as 
\begin{subequations}\label{Bondi variables}
\begin{align}
	F&=2s \kappa+s^2\lambda+\cdots\,,\\
	U^A&=V^A+2s \omega^A+s^2 K^A+\cdots\,,\\
	g_{AB}&=q_{AB}+s C_{AB}+\cdots\,.
\end{align}
\end{subequations}
In the Carter coordinates $(\hv,\hat{s}, \hx^A)$, the near-horizon metric simplifies to 
	\begin{align}\label{NH Carter}
	d{s}^2&=-2 \hat{s} \,\bk\,d\hv^2+2d\hv d\hat{s}-4\hat{s}\,\hat{\omega}_A \,d\hv \, d\hx^A +(\hat{q}_{AB}+\hat{s}\, \hat{C}_{AB})\,d\hx^A d\hx^B+\cO(\hat{s}^2)\,.
\end{align}


\section{Horizon dynamics}\label{sec: horizon dynamics}
\subsection{Kinematic variables on the horizon}
The dynamics of the horizon is best described by a set of intrinsic and extrinsic variables described below. 
Given a coordinate system $(v,x^A)$ on the horizon, and the associated coordinate basis $e_A=\pd_A$, the positive-definite \textit{induced metric} on each cut is defined as
\begin{subequations}\label{kinematic variables}
\begin{align}\label{induced metric def}
	q_{AB} \deq e_A{}^\mu e_B{}^\nu g_\mn\,.
\end{align}
Its determinant is denoted as $q=\det (q_{AB})$. Moreover, the \textit{shift} vector 
\begin{align}\label{velocity field def}
	V^A= \pd_\ell \,x^A
\end{align}
defines a \textit{velocity} field on the horizon. The transverse vector $n^\mu$ is defined such that $\ell\cdot n=-1, n\cdot e_A=0=n\cdot n$, and thus transforms as $n^\mu\to n^\mu/\alpha$ under \eqref{scaling}. The extrinsic variables, including the inaffinity $\kappa$, the Hajicek connection $\omega_A$, and the deformation tensor $\Th_{AB}$ (also called the second fundamental form) are derived from various projections of $\cd_\mu \ell_\nu$ 
\begin{align}\label{NH data}
	\kappa&\deq -\ell^\mu n^\nu  \cd_\mu \ell_\nu\,,\qquad	\omega_A\deq -e_A^\mu n^\nu \cd_\mu \ell_\nu\,,\qquad \Th_{A B}\deq e_A^{(\mu} e_B^{\nu)} \cd_\mu \ell_\nu\,.
\end{align}
The deformation tensor can further be decomposed into the traceless \textit{shear} tensor and the  \textit{expansion} scalar
\begin{align}
	\Theta_{AB}= \sigma_{AB}+\frac12 \th q_{AB}\,,\qquad q^{AB}\sigma_{AB}=0\,.
\end{align}
The geometric variables \eqref{induced metric def},\eqref{velocity field def},\eqref{NH data}, appear directly in the near-horizon expansion of the Carter metric, as evident from \eqref{Bondi variables}. The near-horizon variable $C_{AB}$ turns out to be minus twice the deformation of the transversal vector, \ie  
\begin{align}\label{Cab def}
	C_{AB}=-2e_A{}^{(\mu} e_B{}^{\nu)} \cd_{\mu} n_{\nu}\,.
\end{align}
\end{subequations} 
\subsection{Exact evolution equations}\label{sec: evolution tower}
In this paper, we assume vacuum Einstein equations in a neighborhood of the horizon. For a given coordinate system $(v,x^A)$ on the horizon in which $\ell\deq \pd_v+V^A\pd_A$, the tower of evolution equations on the horizon that we need is given by 
\begin{subequations}\label{evolution tower}
	\begin{align}
		&(\Dv-\th)\sqrt{q}=0\,,\label{evol a}\\
		&(\Dv-\th)q_{AB}=2\sigma_{AB}\,,\label{evol b}\\
		&(\Dv-\kappa)\th=-\frac12 \th^2-\sigma^2\,,\qquad \sigma^2\equiv \sigma_{AB}\sigma^{AB}\label{evol c}\\
		&(\Dv+\th)\omega_A=D_A(\kappa+\th/2)-D_B\sigma_A^B\,,\label{evol d}\\
		&(\Dv-\kappa+\th)\sigma^A{}_{B}=-\cC^A{}_{B}\,,\label{evol e}
	\end{align}
\end{subequations}
where $\Dv$ is the \textit{convective} (fluid) derivative~\cite{Damour:1978cg,Damour:1979wya,Gourgoulhon:2005ng}, and denotes the action of the Lie derivative along  the  generator $\ell\deq \pd_v+V^A\pd_A$. On cut-tensors  $X^{B\cdots}_{A\cdots}(v,x^A)$, it acts as
\begin{align}\label{Dv def}
	\Dv=\frac{d}{dv}+\cL_V\,.
\end{align}
where $\cL_V$ is the two dimensional Lie derivative along $V=V^A\pd_A$.
The first two equations \eqref{evol a},\eqref{evol b} are purely kinematic and follow from $\Th_{AB}=\frac12 \Dv q_{AB}$. 
The third and fourth equations \eqref{evol c},\eqref{evol d} are known respectively as the \textit{null Raychaudhuri} and \textit{Damour-Navier-Stokes} equations, and correspond to $G_{\ell\ell}=G_\mn \ell^\mu \ell^\nu\deq 0,G_{\ell A}=G_\mn \ell^\mu e_A{}^\nu\deq 0$ projections of Einstein equations on the horizon.
Finally, \eqref{evol e} denotes the tidal effects on the horizon due to the ingoing radiation $\cC_{AB}=e_A^\nu e_B^\beta \ell^\mu \ell^\alpha \cC_{\mn\alpha\beta}$, which is the analog of the $\Psi_0$ Weyl scalar in the real basis $e_A$. 
The $G_{AB}\deq 0$ component of Einstein equations, not written here, provides an evolution equation for the transversal deformation tensor $C_{AB}$, whose explicit form is given, \eg in (6.43) of~\cite{Gourgoulhon:2005ng}. 

Note that using \eqref{evol a}, and an integer $p$,
\begin{align}
	q^{-p/2}\Dv \big({q}^{p/2}X \big)=(\Dv+p\,\th)X\,,
\end{align}
using which we can rewrite the second, fourth and fifth equations in \eqref{evolution tower} as 
\begin{subequations}
	\begin{align}
		&\Dv(\sqrt{q}\,q^{AB})=-2\sqrt{q}\,\sigma^{AB}\,,\\
		&\Dv(\sqrt{q}\,\omega_A)=\sqrt{q}\big(\pd_A (\kappa+\th/2)-D_B\sigma_A^B \big)\,,\\
		&(\Dv-\kappa)(\sqrt{q}\,\sigma^A{}_{B})=-\sqrt{q}\,\cC^A{}_{B}\,.
	\end{align} 
\end{subequations}

The operators appearing on the LHS of \eqref{evolution tower} are not accidental. They reflect three covariance properties: 
\begin{enumerate}[leftmargin=0pt]
	\item \textit{Diffeomorphisms of the horizon cuts.} Consider 
	\begin{align}
		v\to v\,,\; x^A\to {x'}^A(v,x)\,.
	\end{align}
	Under this transformation $\ell=\pd_v+V^A\pd_A\to \pd_{v'}+{V'}^A\pd_A$ with 
	\begin{align}
		{V'}^A= \frac{\pd {x'}^A}{\pd x^B}V^B+\pd_v {x'}^A\,,
	\end{align}
	implying that the velocity field transforms as a connection, when $\pd_v {x'}^A\neq 0$. It makes the combination $\Dv=\pd_v+\cL_V$ a covariant derivative under time-dependent diffeomorphisms of the cuts.
	\item  \textit{Rescaling of the generator} $\ell\to \alpha \ell$: for a function that transforms under rescaling as $X\to  \alpha^w X$, the covariant derivative is $(\Dv-w\kappa)$. 
	\item \textit{Weyl transformation of the metric} $g_\mn\to e^{2\lambda}g_\mn$: Under this transformation, $q_{AB}\to e^{2\lambda}q_{AB}$, $\th\to \th+2\Dv\lambda$ and $\kappa\to \kappa+2\Dv\lambda$. For a function that transforms under Weyl rescaling as $X\to e^{p \lambda}X$, covariance is achieved by $(\Dv-\frac{p}{2}\th+q(\th-\kappa))X$ for any $q$. 
\end{enumerate}
If $X$ carries boost weight $w$ and  Weyl
weight $p$, imposing both boost and Weyl covariance fixes $q=w$ and gives
the covariant derivative 
\begin{align}\label{gauge cov evolution}
		\Dv-\frac{p}{2}\theta+w(\theta-\kappa)=\Dv-w\kappa+\left(w-\frac{p}{2}\right)\theta .
\end{align}
For example, the shear $\sigma^A{}_B$ has weights $(w,p)=(1,0)$, and \eqref{gauge cov evolution} reproduces the operator appearing in the LHS of \eqref{evol e}.
Likewise $q_{AB},\sqrt{q}$ have $(w,p)=(0,2)$, so that the covariant derivative $(\Dv-\theta)$ agrees with the LHS of \eqref{evol a},\eqref{evol b}. However, the Raychaudhuri and Damour equations \eqref{evol c},\eqref{evol d} do not transform covariantly under Weyl. This is expected, as they are projections of the Einstein equations which are not Weyl covariant.

\subsection{Multi-scale expansion: quick intro}\label{sec: multiscale intro}

In regular perturbation theory, a function $f(v;\eps)$ with an independent time variable $v$ and a small parameter $\eps$, is perturbed as $f=\sum_{n=0}^N\eps^n f_\po{n}(v)$, with coefficients $f_\po{n}$ independent of $\eps$. In a problem involving several timescales, such an expansion need not
remain uniform on the slow timescale
$v=\cO(\eps^{-1})$, because its coefficients may develop secular
growth. Methods based on scale separation have a long history in gravitational-wave theory. An early example is the short-wavelength expansion introduced by Brill and Hartle and systematized by
Isaacson~\cite{Brill:1964zz,Isaacson:1968hbi,Isaacson:1968zza}. More relevant to our discussion is the two-timescale expansion of extreme mass-ratio inspirals, introduced in~\cite{Hinderer:2008dm}.

An extreme mass-ratio inspiral (EMRI) is a binary system with masses $m,M$ with $\eps\equiv m/M\ll 1$. The dynamics of this system involves at least two timescales: a fast timescale $\tau_\text{orb}\sim\frac{1}{\Omega_\text{orb}}$ given by the orbital period of the secondary object, and a slow \textit{radiation-reaction} timescale $\tau_\text{RR}\sim\frac{\Omega_\text{orb}}{\pd_v\Omega_\text{orb}}$ during which the orbital frequency changes. The ratio is proportional to the perturbation parameter
\begin{align}
\frac{\tau_\text{orb}}{\tau_\text{RR}}=\frac{\pd_v\Omega_\text{orb}}{\Omega_\text{orb}^2}=\cO(\eps)\,.
\end{align}
In the multiple-scale  method, one instead introduces the slow time
$\tlv=\eps v$ and writes the following expansion for a function $f$ on the horizon
\begin{align}
	f(v,x^A;\eps)\sim \sum_{n=0}^N\eps^n f_\po{n}(v,\tlv,x^A)\,,
\end{align}
where the coefficients have no explicit $\eps$ dependence and are
required to remain bounded in the fast variable. We refer the reader to the extensive literature on multiscale analysis of EMRIs~\cite{Hinderer:2008dm,Pound:2010pj,Miller:2020bft,Pound:2021qin}. 
Time derivatives are expanded, using the chain rule, as 
\begin{align}\label{v der}
	\frac{df_\po{n}}{dv}=\pd_v f_\po{n}+\eps \pd_\tlv f_\po{n}\,,
\end{align}
where $v,\tlv$ are viewed as independent variables.  
Consequently, if a coefficient has only slow-time dependence, its
physical time derivative is suppressed by one additional power of
$\eps$. Thus differentiation need not preserve the naive perturbative
order.

\paragraph{Coarse-graining on the horizon.} It is useful to decompose fields on the horizon into an averaged, or
\textit{coarse-grained}, part and an \textit{oscillatory} part. For a scalar function $f$ on the horizon, we define
\begin{align}\label{averaging def}
	\avg{f}(v,\tlv,x)
	&\equiv  \lim_{T\to\infty}\frac{1}{T} \int_0^{T} du\,	f\big(u+v,\tlv,\Phi_u(x)\big),	&	\osc{f}	&\equiv	f-\avg{f}.
\end{align}
where $\Phi_u$ denotes the two dimensional flow by the background velocity field $\bV^A$. 
Said differently, introduce the integral curve $y^A(u,x)$ passing through $x^A$, so that $y^A(0,x)=x^A$ and $\pd_u y^A(u,x)=\bV^A\big(y(u,x)\big)$. Then
\begin{align}
	\avg{f}(v,\tlv,x)	=	\lim_{T\to\infty}	\frac{1}{T}	\int_0^{T}du\,	f\big(u+v,\tlv,y(u,x)\big).
\end{align}
The averaging thus defined satisfies the useful properties
\begin{align}\label{averaging properties}
	\bDv\avg{f}&=0=\avg{\bDv f}\,,& \avg{f\bDv g}&=-\avg{g\bDv f}\,,
	&
	\avg{\avg{f}}&=\avg{f},
	&
	\avg{\osc{f}}&=0,
\end{align}
where $	\bDv\equiv\pd_v+\cL_{\bV}$ is the background convective derivative. 
The averaging \eqref{averaging def} is generalized to a tensor fields $X$ by pulling the tensor back to the reference
point before averaging 
\begin{align}
	\avg{X}\equiv\lim_{T\to\infty}	\frac{1}{T}	\int_0^T du\,	\Phi_u^*X(v+u,\tlv,\Phi_u(x)).
\end{align}

In Carter coordinates $(\hv,\hx^A)$, where $\bV^A=0$, the averaging simplifies to 
\begin{align}\label{averaging Carter}
	\avg{X}(\tlv,\hx^A)
	&\equiv
	\lim_{T\to\infty}
	\frac{1}{T}
	\int_0^T du\,
	X(\hv+u,\tlv,\hx^A)\,,
\end{align}
which agrees with the averaging defined and used in~\cite{Spiers:2026yqx} to address memory effects at null infinity within the multiscale setup; see also \cite{Heisenberg:2023prj}. The averaging \eqref{averaging def} maybe more precisely denoted as $\avg{\cdot}_\bl$, where $\bl=\pd_v+\bV^A\pd_A$ is the fixed background generator whose flow defines the
integration. One could similarly introduce averages along other vector fields, such as the exact horizon generator $\avg{\cdot}_{\ell}$ or the coordinate vector $\avg{\cdot}_{\pd_v}$. In this work we use $\avg{\cdot}_{\bl}$ as averaging along $\ell$ would make the averaging operator field-dependent, whereas averaging along $\pd_v$ is coordinate dependent by construction.

To make the separation into slow and oscillatory sectors explicit, it is useful to use the two-timescale mode expansion of the metric perturbation in an EMRI~\cite{Pound:2021qin}. Considering for simplicity a circular equatorial orbit, a perturbative function on the horizon induced by the bulk perturbation takes the form
\begin{align}\label{mode expansion}
	f(v,\tlv,\theta,\varphi)	=	\sum_{m} f_{m}(\tlv,\theta)	e^{i m (\varphi-\Omega v )}\,.
\end{align}
where $\Omega=\Omega(\tlv)$ is the slowly varying orbital frequency of the secondary. We can now compute the average of \eqref{mode expansion} by recalling that for a Kerr background $\bV =\bW \,\pd_\varphi$ and using the identity
\begin{align}
	\lim_{T\to\infty}\frac{1}{T}\int_0^T du\, e^{im(\bW-\Omega)u}=\de_{m(\bW-\Omega),0}
\end{align}
Evaluating \eqref{averaging def} gives
\begin{align}
	\avg{f}&=\sum_{m} f_{m}(\tlv,\theta) e^{i m(\varphi-\Omega v)}\de_{m(\bW-\Omega),0}
\end{align}
For $\Omega\neq\bW$, only the axisymmetric mode survives $\avg{f}=f_{m=0}(\tlv,\theta)$. On the other hand, if the secondary corotates exactly with the primary black hole, \ie $\Omega= \bW$, all azimuthal modes survive and $\avg{f}=f$. Thus, the coarse-grained sector consists precisely of functions whose
dependence on the fast variables occurs through the corotating angle:
\begin{align}\label{adiabatic functions}
	f=f(\tlv,\th,\varphi-\bW v)\quad \Leftrightarrow\quad f=\avg{f}
\end{align}
For such functions, one can explicitly check that $\bDv f=0$, in accordance with \eqref{averaging properties}.

\subsection{Multiscale dynamics on the horizon}\label{sec: multiscale dynamics}

In this subsection, we analyze the perturbative behavior of the dynamics given by the set of equations \eqref{evolution tower}. We consider the spacetime to be a perturbation of a background stationary and non-extremal black hole, characterized by 
 \begin{align}\label{background data}
 	 \begin{split}
 		q_{AB}=\bq_{AB}(x^A),\quad \omega_A=\bw_A(x^A),\quad \kappa=\bk>0\,,\\
 		\th=0,\quad \sigma_{AB}=0,\quad \cC_{AB}=0,\quad V^A=\bV^A\,.
 	\end{split}
 \end{align}
 Due to axisymmetry $\bV=\bW \,\phi$, where $\phi=\pd_\varphi$ is the axial Killing vector of the Kerr metric, and $\bW$ its constant angular velocity. Stationarity and axisymmetry therefore imply
 \begin{align}\label{bDv Kerr action}
 	\bDv \bV^A=0\,,\qquad \bDv \bw_A=0\,,\qquad \bDv \bq_{AB}=0\,.
 \end{align}
The perturbed geometry can therefore be written up to $\cO(\eps^2)$ corrections as 
\begin{align}\label{perturbation}
			q_{AB}&=\bq_{AB}+\eps q^\po{1}_{AB}\,,&	\kappa&=\bk+\eps \kappa_\po{1}\,,& {V}^A&=\bV^A+\eps V^A_\po{1}\,,&	\omega_A&=\bw_A+\eps \,\omega^\po{1}_A\,,\nonumber \\
		\sigma_{AB}&=\eps \sigma^\po{1}_{AB}\,,& 	\th&=\eps \th_\po{1}\,, & \cC_{AB}&=\eps \cC^\bo{1}_{AB}\,.
\end{align}	
We note that $\kappa, V^A$ parameterize the gauge freedom in the choice of coordinate system $(v,x^A)$ on the horizon. The constraint $\ell(v)=1$ ties this freedom to the scaling of $\ell$ on the horizon. As we will show later in section \ref{sec: coord transf}, fast time dependence in these variables can be gauged away. We therefore impose the following gauge condition
\begin{align}\label{gauge fixing multiscale}
	\osc{\kappa}=0=\osc{V^A}\quad\Leftrightarrow\quad \kappa=\avg{\kappa}\,,\; V^A=\avg{V^A}\quad\Leftrightarrow\quad \bDv\kappa=0=\bDv V^A\,.
\end{align} 
In this subsection, we will study the implications of the evolution equations \eqref{evolution tower} for the geometric quantities introduced above. 

The fluid derivative $\Dv$ can be expanded  as
\begin{align}\label{Dv def linear}
	\Dv=\bDv+\eps \Dv_\po{1}+\cO(\eps^2)\,,\qquad \bDv=\pd_v+\cL_\bV\,,\qquad \Dv_\po{1}= \pd_\tlv+\cL_{V_\po{1}}\,.
\end{align}
Inserting \eqref{perturbation},\eqref{Dv def linear} in the Raychaudhuri equation \eqref{evol c}, and using \eqref{bDv Kerr action}, we find at $\cO(\eps)$
\begin{align}
	(\bDv-\bk)\th_\po{1}&=0\,.
\end{align}
Using the teleological boundary condition that the system settles to a stationary black hole as $v\to \infty$, the exponential solution of the latter equation is excluded and thus $\th_\po{1}=0$. The expansion starts at second order 
\begin{align}\label{th leading}
	\th=\eps^2 \th_\po{2} +\cO(\eps^3)\,,
\end{align}
which according to the Raychaudhuri equation \eqref{evol c} obeys
\begin{align}\label{th2 eq}
	(\bDv-\bk)\th_\po{2}=-\sigma_\po{1}^2\,,\qquad \sigma_\po{1}^2\equiv {\sigma_\po{1}}^A{}_B\,{\sigma_\po{1}}^B{}_A\,.
\end{align}
From this, together with \eqref{averaging properties}, we can infer that $\avg{\th_\po{2}}={\avg{\sigma_\po{1}^2}}/{\bk}$ (see appendix \ref{app: Green} for the exact solution)
The linearized shear $\sigma_\po{1}^A{}_B$ itself can be determined from \eqref{evol e}
\begin{align}
	(\bDv-\bk){\sigma_\po{1}}^A{}_{B}=-{\cC_\po{1}}^A{}_{B}\,,
\end{align}
which implies in particular that $\avg{\sigma_\po{1}^A{}_{B}}={\avg{\cC_\po{1}^A{}_{B}}}/{\bk}$.
Similarly, perturbing  \eqref{evol b} to $\cO(\eps)$ and using \eqref{bDv Kerr action}, \eqref{th leading} gives
\begin{align}
	\bDv q^\po{1}_{AB}+\cL_{V_\po{1}}\bq_{AB}=2\sigma^\po{1}_{AB}\,.
\end{align}
Averaging this equation, using \eqref{averaging properties}, and recalling that by \eqref{gauge fixing multiscale} $V^A_\po{1}=\avg{V^A_\po{1}}$, one finds 
\begin{align}\label{sigma 1 avg}
	\cL_{V_\po{1}}\bq_{AB}=2\bD_{(A}V^\po{1}_{B)}=2\avg{\sigma^{(1)}_{AB}}\,.
\end{align}
Contracting with $\bq^{AB}$ reveals that the linearized velocity field is incompressible 
\begin{align}\label{incompressibility}
	\bD_A V_\po{1}^A=0\,.
\end{align}
Next, perturbing Damour's equation \eqref{evol d} to $\cO(\eps)$, we find 
\begin{align}
	\bDv\omega^\po{1}_A+	\cL_{V_\po{1}}\bw_A=\pd_A\kappa_\po{1}-\bD^B\sigma^\po{1}_{AB}\,.
\end{align}
Averaging implies 
\begin{align}\label{damour averaged}
	 \bD_A\avg{\kappa_\po{1}}-\cL_{V_\po{1}}\bw_A-\bD^B\avg{\sigma^\po{1}_{AB}}=0\,.
\end{align}
Contracting this equation with $V_\po{1}^A$ and integrating over the horizon cut $\cS_v$, one has 
\begin{align}\label{smear V1}
	\int d^2x \sqrt{\bq} V_\po{1}^A \,\left(\bD_A\avg{\kappa_\po{1}}-\cL_{V_\po{1}}\bw_A-\bD^B\avg{\sigma^\po{1}_{AB}}\right)=0\,.
\end{align}
The first term vanishes after integration by parts and using the incompressibility condition \eqref{incompressibility}. For the second term, one uses
\begin{align}
	V_1^A\cL_{V_1}\bw_A=\cL_{V_1}(V_1^A\bw_A)
	=
	V_1^B\bD_B\left(V_1^A\bw_A\right),
\end{align}
whose integral also vanishes by another integration by parts and using the incompressibility condition.
Finally, integration by part on the last term and using \eqref{sigma 1 avg} reduces \eqref{smear V1} to 
\begin{align}
	\int d^2x \sqrt{\bq} \, \avg{\sigma^\bo{1}_{AB}}\avg{\sigma_\po{1}^{AB}}=0\,.
\end{align}
Given the positivity of $\bq_{AB}$, we conclude that the \textit{coarse-grained linearized shear vanishes}
\begin{align}\label{zero avg shear}
	\avg{\sigma^\bo{1}_{AB}}=0\,.
\end{align}
Equation \eqref{sigma 1 avg} therefore reduces to
\begin{align}\label{V1 isometry}
	\cL_{V_\po{1}}\bq_{AB}=0,
\end{align}
showing that $V_\po{1}^A$ is a Killing vector of the background horizon
metric. For a background Kerr black hole, the horizon metric possesses a
single Killing vector, namely the axial Killing vector
$\phi=\pd_\varphi$. We therefore find
\begin{align}\label{V1 Kerr}
	V_\po{1}^A	=	\Omega_\po{1}(\tlv)\phi^A.
\end{align}
Moreover,
\begin{align}
	\cL_{V_\po{1}}\bw_A=\Omega_\po{1}\,\cL_\phi\,\bw_A=0,
\end{align}
due to the axisymmetry of the background. Using these back in the averaged Damour equation \eqref{damour averaged} further
implies
\begin{align}\label{kappa1 multiscale}
	\bD_A\kappa_{(1)}=0 \quad \implies \quad \kappa_{(1)}=\kappa_{(1)}(\tlv).
\end{align}
Thus the slowly varying correction to the inaffinity is purely
monopolar. The same argument goes through for a Schwarzschild background with the small modification: while \eqref{V1 isometry} remains valid, the background $\bq_{AB}$ has three Killing vectors and therefore \eqref{V1 Kerr} is replaced by 
\begin{align}\label{V1 Schd}
	V_\po{1}^A	=	\Omega^{(i)}_\po{1}(\tlv)\phi_{(i)}^A.
\end{align}
which represents a slowly varying rotation about an arbitrary axis. 
Finally, averaging the linearized shear equation and using
\eqref{zero avg shear} gives	$\avg{\cC^{(1)}_{AB}}=0$.
Therefore, within the expansion around a stationary black hole background, the linear shear and the corresponding ingoing Weyl component are purely oscillatory, while the angular velocity $V^A$ is an adiabatic motion along the isometries of the background horizon. Similarly, the slow change $\kappa_\po{1}$ in the inaffinity does not depend on the angles, as its background value $\bk$.  
\paragraph{Remark.} The evolution equations on the horizon do not determine $\kappa_\po{1},\Omega_\po{1}$. These can be inferred by matching this boundary solution to a full bulk solution. This is what we do in section~\ref{sec: coord transf}.

\section{From user gauge to the INU gauge}\label{sec: coord transf}
Consider a perturbation of a nonextremal Kerr spacetime of the form
\begin{align}\label{metric perturbed}
	\tlg^\mn=\bg^\mn+h^\mn\,,
\end{align}
where $h^\mn$ represents the deviation from a background Kerr black hole with inverse metric $\bg^\mn$. We assume that the background metric is already in INU gauge. If this is not the case and the background is written in a different coordinate system (\eg the Boyer-Lindquist), one needs to first apply a finite transformation to put $\bg_\mn$ into the INU form, at least in the vicinity of the horizon~\cite{Fletcher:2003kpg,Scholtz:2017ttf,Kofron:2024taq}. 

We assume that the background is characterized by \eqref{Bondi variables} with
\begin{align}
	\kappa=\bk>0, \quad V^A=\bV^A(x^B)\quad q_{AB}=\bq_{AB}(x^C),\quad \omega_{A}=\bw_A(x^B),\quad C_{AB}=\overline{C}_{AB}(x^C)\,.
\end{align}
where $\bk$ is a constant. For Kerr, $\bV=\bW \phi$ and $\phi=\pd_\varphi$ is a Killing symmetry of $(\bq,\bw,\overline{C})$. 

The deviation $h^\mn$ from the background can be expanded perturbatively in a small parameter $\eps\ll 1$ as 
\begin{align}
	h^\mn=\sum_{n=1}^{N}\eps^n h_\po{n}^\mn
\end{align}
For the purpose of this work, we will restrict to $N=2$, but the procedure can be generalized if needed. The next step is to make a perturbative coordinate transformation~\cite{Bruni:1996im}
\begin{align}\label{knight}
	x^\mu \to	x^\mu-\epsilon \xi_\po{1}^\mu-\epsilon^2\left(\xi_\po{2}^\mu-\frac{1}{2} \xi_\po{1}^\nu \partial_\nu \xi_\po{1}^\mu\right)+O\left(\epsilon^3\right)
\end{align}
under which the metric transforms to ${g}^\mn=\bg^\mn+{\hh}^\mn$ with 
\begin{align}\label{metric gauge transf}
	\hh^\mn= \eps\left(h_1^\mn+\cL_{\xi_\po{1}}\bg^\mn\right)+\eps^2\left(
	h_2^\mn+\cL_{\xi_\po{1}}h_1^\mn+\frac12\cL_{\xi_\po{1}}^2 \bg^\mn+\cL_{\xi_\po{2}}\bg^\mn\right)\,.
\end{align}
$\xi_\po{1},\xi_\po{2}$ are then fixed such that \eqref{INU gauge} is satisfied and the full metric is in INU gauge\footnote{
While we find it easier to work with perturbation of the inverse metric, the problem can equivalently be formulated in terms of the metric perturbation $g_\mn=\bg_\mn+\epsilon \de_1 g_\mn +\eps^2\de_2 g_\mn$. The two are related by 
$h_{1}^\mn=-\bg^{\mu\alpha}\bg^{\nu\beta}\de_1 g_{\alpha\beta}$ and $h_{2}^\mn=\bg^{\mu\alpha}\bg^{\nu\beta}\left(-\de_2 g_{\alpha\beta}+\bg^{\rho\sigma}\de_1 g_{\alpha\rho}{\de_1 g}_{\beta\sigma}\right)$.}.
\subsection{Imposing INU gauge conditions}\label{sec: gauge fixing}
\subsubsection{Bulk gauge fixing} \label{sec:1st order data}
The appropriately gauge transformed perturbation $\hh^\mn=\eps\, \hh_\po{1}^\mn+\eps^2 \,\hh_\po{2}^\mn$ should satisfy the gauge conditions~\eqref{INU gauge}. At linear order
\begin{subequations}\label{NU gauge fixing}
	\begin{align}
		\hh_\po{1}^{vv}&=0:&\pd_s\xi_1^v&=\frac12 h_1^{vv}\\
		\hh_\po{1}^{vs}&=0:&\pd_s\xi_1^s&=h_1^{vs}-\pd_v\xi_1^v-\frac{\bar{F}}{2}h_1^{vv}-\bar{U}^A\pd_A \xi_1^v\\
		\hh_\po{1}^{vA}&=0: & \pd_s\xi_1^A&=h_1^{vA}-\frac12 h_1^{vv} \bar{U}^A-\bar{g}^{AB}\pd_B \xi_1^v
	\end{align}
\end{subequations}
where $\bg^{AB}$ is the inverse of the two dimensional metric $\bg_{AB}$, and for any quantity $X$, we denote by $\bar{X}$ the value of $X$ on the unperturbed Kerr geometry. 
The above equations uniquely determine $\xi^\mu$ in the spacetime in terms of its horizon value 
\begin{align}\label{horizon symmetry}
	\xi_\po{1} &\deq {f}_\po{1}\bl+Q_\po{1} \pd_s+ {Y}_\po{1}^A\pd_A
\end{align}
where $\bl=\pd_v+\bV^A\pd_A$ and $f_\po{1},Q_\po{1},Y_\po{1}^A$ are arbitrary functions of $(v,x^A)$.

\paragraph{Remark.} In practice, we do not need to solve \eqref{NU gauge fixing} exactly. To read off the kinematic variables on the horizon, given by \eqref{kinematic variables}, it is enough to Taylor expand the RHS of \eqref{NU gauge fixing} to second order in $s$ around $s=0$ and solve the resulting equations algebraically. This is usually easier for practical implementations.

\subsubsection{Boundary gauge fixing}
Once the ingoing Newman-Unti gauge \eqref{NU gauge fixing} is fixed, the boundary gauge symmetries \eqref{horizon symmetry} are fixed by imposing three conditions on I) the radial profile of the horizon, II) the velocity field on the horizon, III) the inaffinity on the horizon. 

\begin{enumerate}[leftmargin=0pt]
	\item \textit{horizon locking.}  We can take advantage of the radial component of \eqref{horizon symmetry} to put the perturbed horizon at $s=0$. This requires $\hh^{ss}\deq 0$, which implies at leading order that
	\begin{subequations}\label{horizon gauge eqs}
		\begin{align}\label{Q eq}
			(\bDv-\bk)Q_\po{1}=\frac12 h_\po{1}^{ss}
		\end{align}
		This equation together with the teleological boundary condition for the event horizon, discussed in appendix \ref{app: Green}, completely fix $Q_\po{1}$.
		\item	\textit{Adiabatic velocity. } In INU coordinates, the velocity field reads $V^A\deq g^{sA}$, which is given in terms of user gauge perturbations by 
		\begin{align}\label{Y eq regular}
			V^A&\deq \bV^A+\eps\left(-\Dv Y_1^A+ h_1^{sA}-h_1^{vs}\,\bV^A-(\bD^A-2\bw^A)Q_1\right)+\cO(\eps^2) 			\tag{$\ast$}
		\end{align}
		One is tempted to use $Y_1^A$ to impose the corotation condition $V_\po{1}^A=0$. 
		However, this is not possible in the multiscale setup. Taking the average part of \eqref{Y eq regular}, and noting \eqref{averaging properties}, we observe that the coarse-grained part of $V_\po{1}^A$ cannot be removed. 
		Therefore, instead of imposing the corotation condition, we impose that the velocity field $V^A$ is adiabatic, \ie $V_\po{1}^A=\avg{V_\po{1}^A}$. To this end, we split \eqref{Y eq regular} into its coarse-grained and oscillatory parts. The former is identified with the adiabatic $V_1^A$, while the latter is set to zero by an appropriate choice of $Y_\po{1}^A$
		\begin{align}\label{Y eq multiscale}
			V_1^A&\deq\avg{\cS^A}\,,\qquad \bDv{Y}_1^A\deq\osc{\cS^A}\,,\qquad \cS^A\equiv h_1^{sA}-h_1^{vs}\,\bV{}^A-(\bD^A-2\bw^A)Q_1
		\end{align}		
		\item \textit{Adiabatic inaffinity.} Similarly, the inaffinity is given by $\kappa\deq \frac{1}{2}\pd_s g^{ss}$, which in terms of user gauge perturbation reads
		\begin{align}\label{T eq regular}
			\kappa\deq \bk+\eps\left((\Dv+\bk)(\Dv f_ 1-h_ 1^{vs})+\frac12 \pd_s h_1^{ss}-(2\bw^A\pd_A-\bar{\lambda}) Q_1\right)+\cO(\eps^2)\,.\tag{$\ast\ast$}
		\end{align}
		Following the same philosophy, we split \eqref{T eq regular} into its average and oscillatory parts.  The former fixes $\kappa_\po{1}$, while setting the latter to zero fixes the gauge variable $f_1$
\begin{align}\label{T eq multiscale}
\begin{split}
	&\kappa_\po{1}\deq -\bk\avg{h_1^{vs}}+\frac12\pd_s \avg{h_1^{ss}}+(\bar{\lambda}-2\bw^A \pd_A) \avg{Q_1}\,,\\
	&(\bDv+\bk)\Big(\bDv f_1-\osc{h_ 1^{vs}}\Big)\deq-\frac12 \osc{\pd_s h_1^{ss}}+(2\bw^A\pd_A-\bar{\lambda})\osc{Q_1}\,,
\end{split}
\end{align}
	\end{subequations}
where $\bar{\lambda}$ is background value of $\lambda$ defined in \eqref{Bondi variables}.	
\end{enumerate}

\paragraph{Residual symmetries.} 
The above conditions completely fix $Q_1$, while  $f_1,Y_1^A$ are not completely fixed since the horizon gauge fixing conditions \eqref{Y eq multiscale}, \eqref{T eq multiscale} have homogeneous solutions
\begin{align}\label{homogeneous sol}
(\bDv+\bk)\bDv f_1=0\,,\qquad	\bDv Y_1^A=0\,.
\end{align}
Discarding an exponential solution for $f_1$, the residual symmetries consist of angle-dependent adiabatic functions (see the discussion around \eqref{adiabatic functions} for the exact coordinate dependence of adiabatic functions)
\begin{align}
	(\tl{f}_1,\;\tl{Y}_1^A)\,,\qquad \bDv \tl{f}_1=0=\bDv \tl{Y}_1^A\,.
\end{align}
This freedom does not affect $(\kappa_1,V_1^A)$, but crucially appears at the next order as
\begin{subequations}
	\begin{align}
	V_\po{2}^A&=  -\bDv Y_2^A-\Dv_1 \tl{Y}_1^A+\mathcal{V}_2^A\\
	\kappa_2&=(\bDv+\bk)\bDv f_ 2+\bk\,\Dv_1 \tl{f}_ 1+\mathcal{K}_2
\end{align}
\end{subequations}
where $(\mathcal{V}_2^A,\mathcal{K}_2)$ include second-order perturbative terms, as well as quadratic combinations of first-order perturbative terms.
Splitting these equations into average and oscillatory parts, we find that they can both be set to zero through the gauge symmetries
\begin{align}
	\avg{V_\po{2}^A}=0&\implies \Dv_1 \tl{Y}_1^A=\avg{\mathcal{V}_2^A}\,,&\osc{V_\po{2}^A}=0&\implies \bDv \osc{Y_2^A}=\osc{\mathcal{V}_2^A}\\
	\avg{\kappa_\po{2}}=0&\implies \Dv_1 \tl{f}_1=-\avg{\mathcal{K}_2}/\bk\,,&\osc{\kappa_\po{2}}=0&\implies (\bDv+\bk)\bDv \osc{f_ 2}=-\osc{\mathcal{K}_2}
\end{align}
What is new here, is that the coarse-grained parts can also be set to zero by an appropriate choice of the adiabatic symmetries $(\tl{f}_1,\tl{Y}_1^A)$ induced from the previous perturbative order. This is a possibility that did not exist at first order perturbation. This structure continues to hold at every higher order. We can thus fix the gauge such that  
\begin{align}
	\kappa_{n\geq 2}=0, \qquad V^A_{n\geq 2}=0\,,
\end{align}
so that $\kappa,V^A$ are first-order exact in this restricted gauge
\begin{align}\label{further gauge fixing}
	\kappa=\bk+\eps \kappa_\po{1}(\tlv)\,,\qquad V^A=\bV^A+\eps V_\po{1}^A\,.
\end{align}
Moreover, $\kappa_1,V_1^A$ are dynamically constrained by \eqref{V1 Kerr},\eqref{kappa1 multiscale}. 

Despite the beauty of the extra gauge fixing \eqref{further gauge fixing}, we will not impose it here due to its unpleasant \textit{order mixing} property: imposing $V_2^A=0$ relates $\tl{Y}_1^A$ to the adiabatic part of $V_2^A$. Similarly, imposing $\kappa_2=0$ relates $\tl{f}_1$ to $\kappa_2$. Under $(\tl{f}_1,\tl{Y}_1^A)$, the linearized induced metric and the linearized Hajicek connection transform as 
\begin{align}
	\de_{(\tl{f}_1,\tl{Y}_1)}q_{AB}^\bo{1}&=\cL_{\tl{Y}_1}\bq_{AB},\qquad 
	\de_{(\tl{f}_1,\tl{Y}_1)}\omega^\bo{1}_A=\cL_{\tl{Y}_1} \bw_{A}+\bk \pd_A \tl{f}_1
\end{align}
Therefore, we find that $(q_{AB}^\bo{1},\omega^\bo{1}_A)$ in the restricted gauge \eqref{further gauge fixing} involve $(\kappa_2,V_2^A)$, which in turn depend on $h_2^\mn$ in the user gauge. Similarly, the second order data $(q_{AB}^\bo{2},\omega^\bo{2}_A)$ in the restricted gauge \eqref{further gauge fixing} involve $h_3^\mn$, which takes us beyond the second order perturbation we started with.

\subsection{Perturbed horizon geometry}\label{sec: perturbed horizon geometry}

Equations \eqref{NU gauge fixing} together with \eqref{horizon gauge eqs} determine the gauge transformation required to put the metric in the INU form. Once the metric is in the INU form, the perturbed horizon data can then be computed via \eqref{Bondi variables} from the INU metric
\begin{align}
	V^A\deq g^{sA}\,,\quad q_{AB}\deq g_{AB}\,,\quad \kappa\deq\frac12 \pd_s g^{ss}\,,\quad \omega^{A}\deq\frac12 \pd_s g^{sA}\,,
\end{align}
which can be written explicitly to second order in terms of INU perturbation as
\begin{subequations}\label{linear data INU}
	\begin{align}
		V^A&\deq \bV^A+\eps \,\hh_\po{1}^{sA}+\eps^2 \,\hh_\po{2}^{sA}\,,\\ 
		q_{AB}&\deq \bq_{AB}-\eps\, \hh^\bo{1}_{AB}-\eps^2\, \left(\hh^\bo{2}_{AB}-\bq^{CD} \hh^\bo{1}_{AC}\hh^\bo{1}_{BD}\right)\,,\label{q1 INU}\\
		\kappa&\deq\bk+\frac{\eps}{2} \,\pd_s \hh_\po{1}^{ss}+\frac{\eps^2}{2} \,\pd_s \hh_\po{2}^{ss}\,,\\ 
		\omega^{A}&\deq \bw^A+\frac{\eps}{2} \,\pd_s \hh_\po{1}^{sA}+\frac{\eps^2}{2} \,\pd_s \hh_\po{2}^{sA}\,,
	\end{align}
\end{subequations}
The above equations can be written in terms of the user gauge data perturbatively using \eqref{metric gauge transf}, \eqref{NU gauge fixing}, \eqref{horizon symmetry}. Let us provide the linear data in the following. $V^A_\po{1},\kappa_\po{1}$ are already provided in \eqref{Y eq multiscale},\eqref{T eq multiscale}, which have to take the simple form \eqref{V1 Kerr},\eqref{kappa1 multiscale} as a result of Einstein equations. The induced metric on the horizon is found to be 
\begin{align}\label{induced metric 1st order}
	q_{AB}&=\bq_{AB}+\eps q^\bo{1}_{AB}\,,\qquad q^\bo{1}_{AB}\deq-h^\bo{1}_{AB}+2\bD_{(A}Y^\bo{1}_{B)}+\bar{C}_{AB}Q_1\,.
\end{align}
The deformation tensor can be derived from this through $\Th_{AB}=\frac12\Dv q_{AB}$. Expanding using \eqref{Dv def linear}, together with \eqref{Y eq multiscale}, \eqref{T eq multiscale}, we arrive at
\begin{align}\label{linear deformation}
	\Th^\po{1}_{A B}\deq -\frac12 \bDv h^\bo{1}_{AB}+\bD_{(A} h_\po{1}{}^s{}_{B)}+\frac12\bar{C}_{AB}\bDv Q_\po{1}-\bD_{A}\bD_{B}Q_\po{1}+2\bD_{(A}\big(\bw_{B)} Q_\po{1}\big)\,.
\end{align}
The Hajicek connection is expanded as $\omega^A=\bar{\omega}^A+\eps\,\omega_\po{1}^A$ with
\begin{align}\label{Hajicek 1st order}
	\omega_\po{1}^A\deq\frac12\pd_s h_1^{sA}-\frac12\bD^A h_1^{vs}-\frac12(\bDv +2\bk)h_1^{vA}+\bD^A(\bDv +\bk)f_1+\cL_{Y_1}\bar{\omega}^A+(\bar{K}^A+\frac12 \bar{C}^{AB}\pd_B)Q_1\,.
\end{align}
In these equations, $\bk,\bar{\omega}^A,\bar\lambda,\bar{K}^A,\bar{C}_{AB}$ are the background values of the functions defined in \eqref{Bondi variables}, and spherical indices are moved through the background induced metric $\bq_{AB}$. The functions $Q_\po{1},Y_\po{1}^A,f_\po{1}$ are given by \eqref{horizon gauge eqs}.

\paragraph{Perturbed horizon at second order.}
Once we have fixed $\xi_1^\mu$ in terms of $h_1^\mn$, we can proceed to compute the second order gauge transformation and accordingly the second order data. Recalling \eqref{metric gauge transf}, all the results of section \ref{sec:1st order data} can be generalized to the second order by the replacements 
\begin{align}\label{replacements 2nd order}
	\xi_1\to\xi_2\,,\qquad h_1^\mn\to H_2^\mn= h_2^\mn+\cL_{\xi_\po{1}}h_1^\mn+\frac12\cL_{\xi_\po{1}}^2 \bar{g}^\mn\,.
\end{align}
The second order gauge transformation $\xi_\po{2}^\mu$ is then obtained by the same equations as \eqref{NU gauge fixing}, \eqref{horizon gauge eqs} after the replacements \eqref{replacements 2nd order}. Similarly, the second order perturbed horizon data $\omega_\po{2}^A,\kappa_\po{2},V_\po{2}^A$ can be derived from \eqref{induced metric 1st order},\eqref{Hajicek 1st order} after the replacement \eqref{replacements 2nd order}. However, to derive $q^\bo{2}_{AB}$, we need to additionally add the quadratic term $\bq^{CD} \hh^\bo{1}_{AC}\hh^\bo{1}_{BD}$ in \eqref{q1 INU}. Explicitly,
\begin{subequations}\label{2nd order data}
	\begin{align}
	\kappa_\po{2}&\deq -\bk\avg{H_2^{vs}}+\frac12\pd_s \avg{H_2^{ss}}+(\bar{\lambda}-2\bw^A \pd_A) \avg{Q_2}\,,\\
	V_\po{2}^A&\deq \avg{H_2^{sA}}-\avg{H_2^{vs}}\,\bV{}^A-(\bD^A-2\bw^A)\avg{Q_\po{2}}\,,\\
	q^\bo{2}_{AB}&\deq -H^\bo{2}_{AB}+2\bD_{(A}Y^\bo{2}_{B)}+\bar{C}_{AB}Q_2+\bq^{CD}q^\bo{1}_{AC}q^\bo{1}_{BD}\,,\\
	\omega_\po{2}^A&\deq \frac12\pd_s H_2^{sA}-\frac12\bD^A H_2^{vs}-\frac12(\bDv +2\bk)H_2^{vA}+\bD^A(\bDv +\bk)f_2+\cL_{Y_2}\bar{\omega}^A\nonumber \\
	&\quad +(\bar{K}^A+\frac12 \bar{C}^{AB}\pd_B)Q_2\,.
\end{align}
\end{subequations}
Once we have the induced metric $q_{AB}$ through second order, one can use  
\begin{align}
	\Theta_{AB}= \sigma_{AB}+\frac12 \th q_{AB}=
	\frac12\Dv q_{AB}=\frac12
	\left(\bDv+\eps \Dv_\po{1}+\eps^2 \cL_{V_\po{2}}\right)q_{AB}\,
\end{align}
to obtain the perturbative expressions for the expansion and shear. Using the onshell simplifications \eqref{th leading}, \eqref{V1 isometry}, one finds up to $\cO(\eps^3)$ corrections, that 
\begin{align}
\theta	=\eps^2\theta_2,\qquad 	\sigma_{AB}	=\eps\sigma^\bo{1}_{AB}+\eps^2\sigma^\bo{2}_{AB}\,,	
\end{align}
with
\begin{subequations}\label{2nd order deformation}
\begin{align}
	\sigma^{(1)}_{AB}&=	\frac12\bDv q^{(1)}_{AB},\\
	\theta_2	&=	\frac12\bq^{AB}	\left[	\bDv q^{(2)}_{AB}	+	 \Dv_\po{1} q^{(1)}_{AB}	+	\cL_{V_2}\bq_{AB}	\right]	-	\frac12 q_{(1)}^{AB}\, \bDv q^{(1)}_{AB}\,,\\
	\sigma^{(2)}_{AB}&=	\frac12	\left[	\bDv q^{(2)}_{AB}	+	\Dv_\po{1} q^{(1)}_{AB}	+	\cL_{V_2}\bq_{AB}	-	\th_2\,\bq_{AB}\right].
\end{align}
\end{subequations}
For future reference, note that the Hajicek connection as a covector is expanded as $\omega_A=q_{AB}\omega^B=\bw_A+\eps\, \omega^\bo{1}_A+\eps^2\, \omega^\bo{2}_A$ with 
\begin{align}
	\omega^\bo{1}_A=\bq_{AB}\,\omega_\po{1}^A+q^\bo{1}_{AB}\,\bw^B\,,\qquad \omega^\bo{2}_A=\bq_{AB}\,\omega_\po{2}^A+q^\bo{1}_{AB}\,\omega_\po{1}^B+q^\bo{2}_{AB}\,\bw^B\,,
\end{align}

\subsection{Finite transformation to the Carter coordinates}
\label{sec:finite corotating transformation}

We noted before that a perturbative coordinate transformation of the form \eqref{knight} cannot in general be used to put the metric into the Carter form in a multiscale perturbation. The reason is that perturbative corrections can accumulate into finite $\cO(1)$ shifts over the long timescale. In this subsection, we derive the finite transformation that takes a generic horizon coordinate system $(v,x^A)$, with slowly varying $(\kappa,V^A)$ into the Carter coordinates $(\hv,\hx^A)$ with $\kappa=\bk,V^A=0$.

We first note that equation \eqref{scaling} can be equivalently written as 
\begin{align}\label{lambda eq}
	\ell=\la\, \hl\,,\qquad (\hl+\bk)\la=\kappa
\end{align}
Moreover, we note the definitions $\ell^a\pd_a(x^A)=V^A$ and $\ell^a\pd_a(v)=1$. Rewriting these equations in terms of $\hl$, we find 
\begin{align}
	\hl^a\pd_a(x^A)=\la^{-1}V^A\,,\qquad \hl^a\pd_a(v)=\la^{-1}
\end{align}
In terms of the Carter coordinates $\hx^a=(\hv,\hx^A)$, $\hl=\pd_\hv$. Viewing $\la, V^A$ as functions of the Carter coordinates $\hx^a$, the equations become
\begin{align}\label{old to new}
	\frac{\dd x^A}{\dd \hv}=\la^{-1}V^A\,,\qquad \frac{\dd v}{\dd \hv}=\la^{-1}
\end{align}
From the previous section, the gauge variables are expanded as 
\begin{subequations}\label{alpha expansion}
	\begin{align}
		\kappa&=\bk+\eps\kappa_\po{1}(\cv)+\eps^2\kappa_\po{2}(\cv,\hx^A)+\cO(\eps^3),\\
		V^A &=\bV^A+\eps V_1^A(\cv)+\eps^2 V_2^A(\cv,\hx^A)+\cO(\eps^3)\,,
	\end{align}
\end{subequations}
where $\cv=\eps \hv$ is the adiabatic time of the Carter coordinates. 
To solve  \eqref{lambda eq}, \eqref{old to new}, we consider the following ansatz 
for the map from $x^a$ to $\hx^a$: 
\begin{subequations}\label{INU to Carter}
	\begin{align}
	\la &=1+\eps \la_1(\cv)+\eps^2 \la_2(\cv,\hx^A)+\cO(\eps^3)\\
	v&=\hv+\bsi(\cv)+\eps \psi_1 (\cv,\hx^A)+\eps^2 \psi_2 (\cv,\hx^A)+\cO(\eps^3)\,,\\
	x^A&=\hx^A+\bsi^A(\hv, \cv)+\eps \psi_1^A (\cv,\hx^A)+\eps^2 \psi_2^A (\cv,\hx^A)+\cO(\eps^3)\,.
\end{align}
\end{subequations}
Solving \eqref{lambda eq} specifies the required scaling of the generator
\begin{align}
	\la_1&=\frac{\kappa_1}{\bk}\,, \qquad\la_2=\frac{\kappa_2-\dot{\la}_1}{\bk}	
\end{align}
while solving \eqref{old to new} gives the explicit transformation between the two coordinates
\begin{subequations}\label{coord transf}
	\begin{align}
	\bsi&=-\int^\cv du \la_1(u)\,,& \psi_1&=-\int^\cv du \left(\la_2-\la_1^2\right)  \label{v transf} \\
	\bsi^A&=\hv \bV^A +\int^\cv du \left(V_1^A-\la_1\bV^A\right)\,,& \psi_1^A&=\int^\cv du \left(V_2^A-\la_1 V_1^A-(\la_2-\la_1^2)\bV^A\right)\,.\label{xA transf}
\end{align}
\end{subequations}

\paragraph{Transformed basis.} 
We note from \eqref{v transf} that the $v$-slicing of the horizon coincides with $\hx$-slicing at $\cO(\eps^0)$. However, this is not the case when subleading transformation is taken into account. Therefore, the basis vector $\he_A=\frac{\pd}{\pd \hx^A}\big\vert_\hv$ cannot be the same as $e_A=\pd_A\big\vert_v$. Using the chain rule
\begin{align}
	\he_A=\frac{\pd v}{\pd \hx^A}\pd_v+\frac{\pd x^B}{\pd \hx^A}\pd_B=\frac{\pd v}{\pd \hx^A} \ell+\left(\frac{\pd x^B}{\pd \hx^A}-V^B \frac{\pd v}{\pd \hx^A}\right)\pd_B
\end{align}
where $\ell=\pd_v+V^A\pd_A$. The basis transformation can therefore be written as 
\begin{align}\label{new basis}
	\he_A=E_A{}^B \left(e_B+\beta_B \ell\right)\,,\qquad E_A{}^B=\frac{\pd x^B}{\pd \hx^A}-V^B \frac{\pd v}{\pd \hx^A}\,,\qquad \beta_B=(E^{-1})^A{}_B \,\frac{\pd v}{\pd \hx^A}
\end{align}
The transversal null vector $\hn$ is then exactly fixed by the orthogonality conditions. The Carter frame is then
\begin{align}\label{tetrad transform}
	\hl=\frac{1}{\la}\ell\,,\qquad \he_A=E_A{}^B \left(e_B+\beta_B \ell\right)\,,\qquad \hn=\la \left(n+\frac12 \beta^2\ell+\beta^A e_A\right)
\end{align}
where $\beta^A=q^{AB}\beta_B$, $\beta^2=\beta^A\beta_A$. Eq. \eqref{tetrad transform} corresponds to the composition of type I and type III null rotations in the language of Newman-Penrose~\cite{Chandrasekhar:1985kt}. 

Using \eqref{new basis} and the perturbative coordinate transformation \eqref{coord transf}, we can find the perturbative expansion of $E_A{}^B, \beta_A$ as 
\begin{subequations}\label{new basis perturbative}
	\begin{align}
	E_A{}^B&=\de_A{}^B+\eps\left(\pd_A \psi_1^B-\bV^B\pd_A \psi_1\right)+\eps^2\left(\pd_A \psi_2^B-V_1^B\pd_A \psi_1-\bV^B\pd_A \psi_2\right)\,,\\
	\beta_A&=\eps \pd_A \psi_1+\eps^2 \left(\pd_A\psi_2-(\pd_A \psi_1^B-\bV^B\pd_A \psi_1)\pd_B \psi_1\right)
\end{align}
\end{subequations}
where it is understood that $\pd_A=\frac{\pd}{\pd\hx^A}$.
\paragraph{Transformed horizon data.}
The transformed horizon data in Carter coordinates is then found as 
\begin{subequations}\label{transformed data}
	\begin{align}
	\hat{q}_{AB}&=E_A{}^C E_B{}^D q_{CD}\,\\ 
	\hat{\Th}_{AB}&=E_A{}^C E_B{}^D \la^{-1} \Th_{CD}\\
	\hat{\omega}_A&=E_A{}^B\left(\omega_{B}-D_B\ln \la +\bk\la \beta_B -\Th_{BC}\beta^C\right)
\end{align}
\end{subequations}The perturbative expansion of the horizon data in Carter coordinates is therefore obtained by \eqref{transformed data} and the perturbative expansions \eqref{new basis perturbative} and the results found in section \ref{sec: perturbed horizon geometry}.

The horizon evolution equations take a particularly simple form in Carter coordinates. By construction $
	\hl=\pd_{\hv},
	\hat{\kappa}=\bk,
	\hat{V}^A=0$
and hence the convective derivative reduces to
$\hat{\Dv}=\dd_{\hv}=\ddhv$. Therefore, the evolution equations
\eqref{evolution tower} become
\begin{subequations}\label{evolution tower Carter}
	\begin{align}
		&(\dd_{\hv}-\hth)\sqrt{\hat q}
		=0,
		\label{evol a'}\\
		&(\dd_{\hv}-\hth)\hat q_{AB}
		=2\hat\sigma_{AB},
		\label{evol b'}\\
		&\left(\dd_{\hv}-\bk\right)\hth
		=
		-\frac12\hth^2-\hat\sigma_{AB}\hat\sigma^{AB},
		\label{evol c'}\\
		&(\dd_{\hv}+\hth)\hat\omega_A
		=
		\frac12\hat D_A\hth
		-\hat D_B\hat\sigma_A{}^B,
		\label{evol d'}\\
		&(\dd_{\hv}-\bk+\hth)
		\hat\sigma^A{}_B
		=
		-\hat{\cC}^A{}_B.
		\label{evol e'}
	\end{align}
\end{subequations}
Here $\hat D_A$ is the covariant derivative associated with
$\hat q_{AB}$, and all indices in \eqref{evolution tower Carter} are
raised and lowered using $\hat q_{AB}$.

\paragraph{Order mixing.}
One subtlety of the Carter coordinate system is that the perturbative
expressions for $\hat q_{AB}$ and $\hat\omega_A$ through second order
require $E_A{}^B$ through the same order, as is clear from
\eqref{transformed data}. According to \eqref{new basis perturbative},
this in turn requires $(\psi_2,\psi_2^A)$. In \eqref{coord transf}, we observed that to derive $\psi_1^a=(\psi_1,\psi_1^A)$, one needs to solve \eqref{old to new} to second order. To find  $\psi_2^a=(\psi_2,\psi_2^A)$, we need to solve  \eqref{old to new} to third order, which requires the knowledge of $(\kappa_3,V_3^A)$. Thus, although the horizon data in the original
INU frame are determined by a second-order bulk perturbation, imposing
the Carter conditions through second order requires partial information
from the third-order perturbation. We call this the \textit{order mixing} problem, which also existed in the restricted gauge \eqref{further gauge fixing}.

We can extend \eqref{coord transf} to the next order to find $\psi_2^a$. It is sourced by third-order perturbation $\kappa_3,V_3^A$, as well as nonlinear terms due to nonlinear lower order perturbation. Since the third order data are not known, $\psi_2^a$ is ambiguous. The resulting ambiguity has, however, a restricted form. The difference between two third-order completions $\psi_2^a,\tl{\psi}_2^a$---with the former including the effect of $\kappa_3,V_3^A$, and the latter ignoring their contribution---can be regarded as an infinitesimal diffeomorphism under 
\begin{align}\label{order mixing residual}
	\hat x^a\longrightarrow
	\hat x^a-\eps^2\zeta^a+\cO(\eps^3),
	\qquad
	\zeta=F\,\pd_{\hat v}+Z^A\pd_A .
\end{align}
where 
\begin{align}
	F&=\psi_2-\tl{\psi}_2=-\frac{1}{\bk}\int^\cv du \,\kappa_3\,,\qquad Z^A=\psi^A_2-\tl{\psi}^A_2=\int^\cv du\, \left(V_3^A-\frac{\kappa_3}{\bk}\bV^A\right)
\end{align}
Under this ambiguity, 
\begin{align}\label{order mixing data}
	\Delta_\zeta q^{(2)}_{AB}
	&=\cL_Z\bar q_{AB},
	\qquad
	\Delta_\zeta\omega^{(2)}_A
	=\cL_Z\bar\omega_A
	+\bk \bD_A F\,,\qquad \Delta_\zeta \Th_{AB}=\cO(\eps^3)\,.
\end{align}
The last identity follows from the fact that $\Th_{AB}$ vanishes on the background. 

General horizon charges, constructed in Carter coordinates, are sensitive to the shifts \eqref{order mixing data} and therefore cannot, in general, be reconstructed through second order from a second-order bulk perturbation. Importantly, this ambiguity drops out of the global horizon charges relevant below. In the next section we show explicitly that energy, angular momentum and (dynamical) entropy are invariant under \eqref{order mixing data}. 

\section{Horizon charges and fluxes}\label{sec:charges}

\subsection{Carter coordinates and its residual symmetries}
In section \ref{sec: coord transf}, we explained how to put a generic two-timescale perturbation of a background Kerr black hole into the ingoing Newman-Unti form and read off the horizon data. Furthermore, we discussed in section \ref{sec:finite corotating transformation} that to transform the metric into the Carter gauge, one needs to perform an additional \textit{finite} coordinate transformation. In this section, we assume that these two steps are taken and the horizon data are found in the Carter gauge. Our goal is now to discuss black hole charges and their fluxes through the horizon, using the associated covariant phase space.

\paragraph{Notational remark.} All of the discussion of this section is performed in the Carter coordinates $(\hv,\hx^A)$. In previous sections, we used hatted quantities to denote horizon data in Carter coordinates. To avoid notational clutter, we will drop all the hats, except when referring explicitly to the coordinates.

Residual symmetries of the Carter coordinates $(\hv,\hx^A)$, in which the metric takes the form \eqref{NH Carter}, are given by vector fields of the form
\begin{align}\label{xi def}
	\xi\deq f\pd_\hv+Y^A \pd_A
\end{align}
subject to 
\begin{align}\label{symmetry constraints}
	(\pd_\hv+\bk)\pd_\hv f=0\,,\qquad \pd_\hv Y^A=0
\end{align}
We note that the symmetries are simply the homogeneous solutions to the gauge fixing conditions \eqref{homogeneous sol} written in Carter coordinates. The former equation can be solved as 
\begin{align}\label{f decomp}
	f=T(\hx^A)+W(\hx^A)e^{-\bk \hv}
\end{align} 
The symmetry vector field thus expands as 
\begin{align}\label{symmetries}
	\xi\deq \Big[T(\hx^A)+W(\hx^A)e^{-\bk \hv}\Big]\pd_\hv+Y^A(\hx^B)\pd_A
\end{align}
By analogy with the terminology of null infinity, $T(\hx^A) \pd_\hv$ may be called a supertranslation, and  $Y^A(\hx^B)\pd_A$ a superrotation\cite{Donnay:2019jiz}.
However, recall that we have used a Killing parametrization of the background $\hell\cdot \cd \hell=\bk \hell$, while the generator of null infinity is affinely parametrized $n\cdot \cd n=0$. We can rescale $\hell_\af \equiv e^{-\bk \hv}\hell$, so that $\hell_\af\cdot \cd \hell_\af=0$. This corresponds to using an affine parameter $\tau$ along the horizon generator, related to the Killing parameter $\hv$ by
\begin{align}\label{affine vs Killing}
	\tau=\frac{e^{\bk \hv}}{\bk},
	\qquad
	\tau\pd_\tau=\frac{1}{\bk}\pd_\hv .
\end{align}
In terms of the affine parameter, the symmetry generator reads
\begin{align}\label{symmetries affine}
	\xi\deq \Big[W(\hx^A)+\bk \,\tau \,T(\hx^A)\Big]\pd_\tau+Y^\hA(\hx^B)\pd_\hA
\end{align}
In this parametrization, which makes the analogy to null infinity more direct, it is more natural to identify $W$ with a supertranslation, while $T$ is an angle-dependent dilation~\cite{Ashtekar:2021kqj,Ashtekar:2021wld}. Nonetheless, analogy with null infinity will not be necessary for what follows.

\subsection{Charges and fluxes from the covariant phase space}

Using the Carter coordinates, we now have an infinite-dimensional class of geometries close to the Kerr
black hole of the form \eqref{NH Carter}, with symmetry generator \eqref{symmetries}. Using the notation $\cL_\xi X\equiv (f\frac{\dd}{\dd\hv} +\cL_\cY)X$ for a tensor intrinsic to the cuts, horizon data is transformed under symmetries as $\de_\xi\kappa=0=\de_\xi V^A$, 
\begin{subequations}
	\begin{align}
		\de_{\xi}\, q_{AB}&=\cL_\xi q_{AB}\,,\\
		\de_{\xi} \,\Th_{AB}&=\cL_\xi \Th_{AB}+\frac{\dd}{\dd\hv} f \,\Th_{AB}\,,
	\end{align}
while 
\begin{align}
	\de_{\xi} \,\omega_{A}\;&= \cL_\xi \omega_{A}+\Delta_f\omega_A\,,\qquad \Delta_f\omega_A\equiv (\tfrac{\dd}{\dd\hv}+\bk) D_A f-\Th_{A}{}^B\pd_B f 
\end{align}
\end{subequations}
contain an anomalous term. This solution space can be endowed with a presymplectic structure on the horizon, from which one can construct
charges and flux-balance laws associated with the residual symmetries
\eqref{symmetries}. The presymplectic potential so defined is not unique: it may be shifted by total derivative and total variation terms. Different prescriptions  have been used in the literature to satisfy various boundary conditions~\cite{Chandrasekaran:2018aop,Aghapour:2018icu,
	Jafari:2019bpw,Donnay:2019jiz,Adami:2021nnf,Ashtekar:2021kqj,
	Ashtekar:2021wld,Ciambelli:2023mir,Hopfmuller:2018fni,Ashtekar:2024stm}. 
	
In this work, we will follow~\cite{Chandrasekaran:2018aop} whose phase space is constructed by considering the equivalence class $[\ell^\mu,\kappa]$ of horizon generator and inaffinity (related by \eqref{scaling}) as fixed structures
\begin{align}\label{CFP gauge conditions}
	\delta[\ell^\mu]\deq0,
	\qquad
	\delta[\kappa]\deq0\,.
\end{align}
In the Carter coordinate system,  $\kappa=\bar\kappa$, and $\ell\deq \ddhv$ and hence their constraints are exactly obeyed. Therefore, we can use their phase space construction. To uniquely determine charges and fluxes, they apply the Wald--Zoupas prescription \cite{Wald:1999wa}, requiring the boundary presymplectic potential to vanish whenever the null hypersurface is shear-free and expansion-free. The resulting charge corresponding to symmetry generator \eqref{xi def}, subject to \eqref{symmetry constraints}, is 
\begin{align}\label{charge def full}
	Q_\xi&= \frac{1}{2}\oint \sqrt{q} \,\left[f\big(\bk-\th \big)+\pd_\hv f+Y^A\omega_A\right]\,,
\end{align}
which can be split into $Q_f$ corresponding to $\xi=f\pd_\hv$ and $Q_Y$ corresponding to $\xi=Y^A\pd_A$\footnote{Note that $Q_f$ here differs by a sign to that of \cite{Chandrasekaran:2018aop} so that we get the correct sign for entropy. A more general analysis of \cite{Odak:2023pga} identifies a one-parameter family of covariant null symplectic potentials in which the supertranslation charge takes the form $	Q^{(c)}_{\mathcal T}=\frac12\oint\sqrt q\,\mathcal T\left(\bar\kappa-\frac{c}{2}\theta\right)$. The Wald--Zoupas stationarity requirement is satisfied in the general case if the variations of the inaffinity and the expansion are constrained according to $\delta\kappa=\frac{c-2}{2}\,\delta\theta$. In the Carter gauge $\de \kappa=0$, while $\delta\theta$ is nonzero, which naturally selects $c=2$ corresponding to the choice of \cite{Chandrasekaran:2018aop}.} 
\begin{align}\label{charges def}
	Q_f&= \frac{1}{2}\oint \sqrt{q} \,\left[f\big(\bk-\th \big)+\pd_\hv f\right] \,,\qquad Q_Y= \frac{1}{2}\oint \sqrt{q} \,Y^A\omega_A\,.
\end{align}
where we use the shorthand notation $\oint\equiv \int\frac{d^2\hx}{4\pi}$.
The former charge can be decomposed, using  \eqref{f decomp}, as 
\begin{align}\label{f charge split}
	Q_T&=\frac{1}{2}\oint \sqrt{q} \,T\,\big(\bk-\th \big) \,,\qquad Q_W=-\frac12 \oint \sqrt{q} \,W \,\th_\af\,.
\end{align}
where we have used the discussion around \eqref{affine vs Killing} to identify $\th\,e^{-\bk\hv}$ with the expansion of the horizon in the affine parametrization  $\th_\af$.
The charges \eqref{charges def} obey certain balance equations of the form 
\begin{align}
	\dd_\hv Q_\xi=\cF_\xi
\end{align}
 that specifies their evolution along the horizon. Applying $\dd_\hv$ on \eqref{charges def} and using the evolution equations \eqref{evolution tower Carter}, one finds that the flux corresponding to the generator $f\pd_\hv$ is 
\begin{subequations}\label{T flux}
	\begin{align}
		\cF_f&=\frac12\oint \sqrt{q} \,f\Big(\sigma_{AB}\sigma^{AB}-\frac12 \th^2\Big)\\
		&=\frac14 \oint \sqrt{q}\,f\Big(\sigma^{AB}- \frac12\th q^{AB}\Big)\dd_\hv q_{AB}
	\end{align}
\end{subequations}
where we used \eqref{evol a'},\eqref{evol b'}, and that $\dd_\hv \ln \sqrt{q}=\frac12 q^{AB}\dd_\hv q_{AB}$. For superrotation fluxes, corresponding to the generator $Y^A\pd_A$, we find
\begin{subequations}\label{Y flux}
	\begin{align}
	\cF_Y&=\frac14 \oint \sqrt{q}\Big(\sigma^{AB}\cL_Y q_{AB}- \th D_A Y^A\Big)\\
	&=\frac14 \oint \sqrt{q}\Big(\sigma^{AB}- \frac12\th q^{AB}\Big)\cL_Y q_{AB}\label{QY flux}
\end{align}
\end{subequations}
The second lines of \eqref{T flux}, \eqref{Y flux} can be unified into a single flux formula for the general symmetry transformation $\xi\deq f\pd_\hv +Y^A \pd_A$ as
\begin{align}\label{flux v1}
	\boxed{\cF_{\xi}=\frac14 \oint \sqrt{q}\,\Big(\sigma^{AB}- \frac12\th q^{AB}\Big)\de_\xi q_{AB}}
\end{align}
where $\de_\xi q_{AB}\equiv f\frac{\dd}{\dd \hv} q_{AB}+\cL_Y q_{AB}$. Note that $\sigma^{AB}- \frac12\th q^{AB}=\Th^{AB}-\Th q^{AB}$ is the trace-reversed deformation tensor.
This flux formula agrees with (4.13) of~\cite{Ashtekar:2021kqj} and is a nonperturbative extension of the perturbative flux in \cite{Poisson:2004cw}. The flux \eqref{flux v1} can be equivalently written as either of
\begin{align}\label{flux alternative}
	\cF_{\xi}	&=	-\frac18	\oint	\frac{1}{\sqrt q}\,	\dd_{\hat v}\!\left(q q^{AB}\right)	\mathcal L_\xi q_{AB}
	=-\frac18	\oint	q^{3/2}	\dd_{\hat v}q^{AB}\,	\mathcal L_\xi\!\left(\frac{q_{AB}}{q}\right).
\end{align}
The powers of $q$ suggest splitting of the induced metric into the area density $\mu$, and its unimodular conformal \textit{shape} $\gamma_{AB}$
\begin{align}
	q_{AB}=\mu \,\gamma_{AB}\,,\qquad \mu=\sqrt{q}\,,\qquad \det\gamma_{AB}=1
\end{align}
Here $\mu$ and $\gamma_{AB}$ are tensor densities of weights $1$ and $-1$, respectively, and their Lie derivatives below are understood as density-weighted Lie derivatives.
In terms of these variables, \eqref{evolution tower Carter} give
\begin{align}\label{evol shape}
	\dd_\hv \mu=\mu\,\hth\,,\qquad \dd_\hv \gamma_{AB}=\frac{2\sigma_{AB}}{\mu}\,,\qquad \dd_\hv \gamma^{AB}=-2\mu\,\sigma^{AB}
\end{align}
where $\gamma^{AB}$ is the inverse of $\gamma_{AB}$. Accordingly, the flux formula \eqref{flux alternative} is rewritten, using \eqref{evol shape}, as
\begin{align}\label{flux shape area}
	\cF_{\xi}
	&=\frac{1}{4}\oint \left(\mu^2 \, \sigma^{AB} \cL_\xi \gamma_{AB}- \hth \cL_\xi \mu\right)
\end{align}
which clearly shows the pairing between the radiative modes $(\gamma_{AB},\sigma^{AB})$ on the one hand and $(\hth,\mu)$ on the other hand.
\subsection{Global charges}
In this subsection, we will introduce a finite dimensional subset of the charges \eqref{charges def} that are of more interest to us. All of these charges have the property that they are conserved at leading order perturbation, \ie their fluxes are $\cO(\eps^2)$. In particular, we will focus on I) dynamical entropy, II) Lorentz charges, III) energy.

\paragraph{Dynamical entropy.} 
The charge and flux corresponding to $T=1$ are
\begin{align}\label{Q T=1}
	Q_{T=1}
	&=\frac{1}{8\pi}\left(\bk A-\dd_\hv A\right)\,,
	\qquad
	\dd_\hv Q_{T=1}
	=\frac12\oint \sqrt{q}\,
	\left(\sigma_{AB}\sigma^{AB}-\frac12\th^2\right),
\end{align}
where $	A(\hv)=\int_{\cS_{\hv}}d^2\hx\,\sqrt{q}$ is the area of the horizon cut $\cS_{\hv}$. This charge admits a direct
thermodynamic interpretation also away from stationarity. Recall the relation between affine and Killing parametrization of the horizon \eqref{affine vs Killing}. 
For perturbations of a stationary Killing horizon, the dynamical entropy
introduced in \cite{Hollands:2024vbe,Visser:2024pwz} is
\begin{align}\label{dynamical entropy}
	S_{\rm dyn}
	&=\left(1-\tau\pd_\tau\right)S,\qquad S=\frac{A}{4}\,.
\end{align}
It therefore follows that
\begin{align}\label{Sdyn def}
		Q_{T=1}=T_{\rm H}S_{\rm dyn},
	\qquad
	T_{\rm H}=\frac{\bk}{2\pi}.
\end{align}
Thus, the term $-\dd_\hv A$ in the charge \eqref{Q T=1} is precisely the dynamical
correction to the Bekenstein--Hawking entropy, which vanishes in a stationary era. The balance equation \eqref{Q T=1} can equivalently be written as an entropy-production law
\begin{align}\label{Sdyn flux}
	\ddhv S_{\rm dyn}
	=
	\frac{1}{4\bk}
	\int_{\cS_{\hv}}d^2\hx\,\sqrt{q}\,
	\left(\sigma_{AB}\sigma^{AB}-\frac12\th^2\right).
\end{align}
In the perturbative expansion considered here,
$\sigma_{AB}=\cO(\eps)$ while $\th=\cO(\eps^2)$. Hence the leading
nontrivial entropy production is
\begin{align}
	\ddhv S_{\rm dyn}
	=
	\frac{\eps^2}{4\bk}
	\int_{\cS_{\hv}}d^2\hx\,\sqrt{\bq}\,
	\sigma^\bo{1}_{AB}\sigma_\po{1}^{AB}
	+\cO(\eps^3),
\end{align}
which is manifestly non-negative. 
\paragraph{Lorentz charges.}
	As a two-dimensional metric, $\bq_{AB}$ necessarily has six independent conformal Killing vectors (CKV) $\bY^A$ such that 
	\begin{align}\label{CKV}
		\cL_{\bY} \bq_{AB}=\bsi \,\bq_{AB}\,,\qquad  \bsi=\bD_A \bY^A
	\end{align}
At most, three of these can be a Killing vector ($\cL_Y \bq_{AB}=0$). In a Schwarzschild geometry, these Killing vectors are induced from the rotational symmetries of the spacetime. For an axisymmetric geometry like Kerr, only one Killing vector $\phi=\pd_\varphi$ is guaranteed. 
A Lorentz transformation is not a pure CKV of the cut. In analogy with null infinity, we identify Lorentz transformations with $\xi\deq Y^A\pd_A+\frac12\psi_Y \tau\pd_\tau$,
where $\tau$ is an affine parameter on the horizon generators~\cite{Ashtekar:2021wld,Ashtekar:2021kqj}. Recalling \eqref{affine vs Killing}, a Lorentz transformation is generated by 
\begin{align}\label{Lorentz generator}
	\chi\deq \bY^A\pd_A+\frac{\bsi}{2\bk} \pd_\hv\,,\qquad \bY^A=\text{CKV of } \bq_{AB}
\end{align}
and the corresponding charge is
\begin{align}
	Q_\chi&=Q_\bY+\frac{1}{2\bk}Q_{T=\bsi}
\end{align}
In the special case where $\bY=-\phi$ is a Killing vector, $\bsi=0$ and $Q_\chi=Q_\bY$. In particular, for the rotational symmetry of the Kerr black hole,  $\phi=\pd_\varphi$, the associated charge defines the angular momentum of the black hole (the extra minus sign is standard, see \eg \cite{Wald:1993nt})
\begin{align}\label{J def}
	J&=-Q_{\phi}
\end{align}

The Lorentz fluxes can be written starting from \eqref{flux shape area}. First, note that \eqref{CKV} implies $\cL_\bY \bgamma_{AB}=0$, \ie that $\bY^A$ is an isometry of the conformal shape tensor density. This readily implies that the Lorentz fluxes are at least $\cO(\eps^2)$. Expanding
\begin{align}
	\mu&= \bar\mu \big(1+\eps\mu_1+\cO(\eps^2)\big)\,,\qquad
	\gamma_{AB}=\bgamma_{AB}+\eps \gamma^\bo{1}_{AB}+\eps^2 \gamma^\bo{2}_{AB}+\cO(\eps^3)
\end{align}
we find
\begin{align}\label{Lorenz flux 2ndO}
	\cF_{\chi}&=\frac{\eps^2}{4}\oint \bar{\mu} \left[\bar{\mu}\sigma_\po{1}^{AB}\cL_\bY \gamma^\bo{1}_{AB}-\bsi \left(\th_\po{2}-\sigma_\po{1}^2/\bk\right)\right]
\end{align}
Interestingly, the time component of the Lorentz vector leads to the $\sigma_\po{1}^2$ term in the last bracket. Using the Raychaudhuri equation, the last parentheses combine into $\frac{1}{\bk}\pd_\hv \hth_\po{2}$, which is an oscillatory variable and averages to zero. Therefore, the coarse-grained Lorentz flux simplifies considerably
\begin{align}
	\avg{\cF_{\chi}}&=\frac{\eps^2}{4}\oint \bar{\mu}^2\avg{\sigma_\po{1}^{AB}\cL_\bY \gamma^\bo{1}_{AB}}
\end{align}

\paragraph{Energy.}
Defining energy is surprisingly more subtle than the other charges in
this phase space. A first guess would be to define it as the charge
corresponding to the time-evolution vector field of the background Kerr
solution,
\begin{align}\label{t vs v}
	\bt=\pd_\hv-\bW\,\phi\,.
\end{align}
However, inserting this into the charge expression \eqref{charges def}
and evaluating it on the background solution gives
\begin{align}
	Q_\bt=\frac{\bk}{8\pi}A+\bW J\,.
\end{align}
Using the Smarr formula for Kerr,
$M_\kerr=\frac{\bk}{4\pi}A+2\bW J$, we observe that
$Q_\bt=M_\kerr/2$, which is off by a factor of two. Adding by hand half
of the background mass is not a resolution, because we want to
reproduce the energy on the whole Kerr family, rather than only on the
chosen reference solution. We therefore follow a different path.

On the Kerr family, viewed as a subspace of the dynamical horizon phase
space, the energy is given by the Kerr mass. In terms of the
Bekenstein--Hawking entropy $S=S_{\text{BH}}=A/4$ and angular momentum $J$, it can be
written as
\begin{align}\label{Kerr mass}
	M_\kerr(S,J)
	=
	\sqrt{\frac{S}{4\pi}+\frac{\pi J^2}{S}}\,.
\end{align}
The corresponding surface gravity $\kappa_\kerr{\scriptstyle{(S_\dyn,J)}}$ and angular velocity $\Omega_\kerr{\scriptstyle{(S_\dyn,J)}}$ satisfy
\begin{align}\label{Kerr properties}
	\frac{\kappa_\kerr}{2\pi}
	&=
	\frac{\pd M_\kerr}{\pd S}\,,
	&
	\Omega_\kerr
	&=
	\frac{\pd M_\kerr}{\pd J}\,,
	&
	\frac{\pd\kappa_\kerr}{\pd J}
	&=
	2\pi\frac{\pd\Omega_\kerr}{\pd S}\,.
\end{align}
They are explicitly
\begin{align}\label{Kerr potentials}
	\kappa_\kerr
	&=
	\frac{1-(2\pi J/S)^2}
	{2\sqrt{S/\pi}\sqrt{1+(2\pi J/S)^2}}\,,
	&
	\Omega_\kerr
	&=
	\frac{2\pi J/S}
	{\sqrt{S/\pi}\sqrt{1+(2\pi J/S)^2}}\,.
\end{align}
The Kerr mass obeys the Smarr relation
\begin{align}\label{Smarr}
	M_\kerr
	=
	2\left(
	\frac{\kappa_\kerr}{2\pi}S+\Omega_\kerr J
	\right),
\end{align}
while its variation gives, upon using \eqref{Kerr properties} the first law,
\begin{align}
	\de M_\kerr
	=
	\frac{\kappa_\kerr}{2\pi}\de S
	+\Omega_\kerr\de J\,.
\end{align}

We now want to extend this definition of energy off the Kerr family.
Such an extension is not unique. We propose the following extension due
to its simple properties. First note that on the Kerr family,
$S_\dyn=S=A/4$. For a generic dynamical horizon,
$S_\dyn$ and $J$ can be unambiguously computed using
\eqref{charges def} with $T=2\pi/\bk$ and $Y=-\phi$, respectively.
We then define the energy associated with a dynamical black hole by
retaining the Kerr equation of state and replacing $S$ by
$S_\dyn$:
\begin{align}\label{Energy def}
		E 		\equiv
		M_\kerr(S_\dyn,J)\,.
\end{align}
Equivalently, using the Smarr relation\footnote{The energy \eqref{Energy Smarr} coincides with the charge $Q_t$ computed through \eqref{charge def full} with symmetry parameter $t=2(\frac{\kappa_\kerr}{\bk}\pd_\hv-\Omega_\kerr \phi)$. However, the vector $t$ is \textit{field-dependent} and the identification of charges and fluxes for field dependent symmetries requires careful investigation. We leave this issue, \ie a first-principle definition of the energy on a dynamical horizon, for a future work. The need for field-dependent symmetries to identify energy has been stressed in~\cite{Ashtekar:2021kqj}. The covariant phase space with field dependent symmetries is developed in \cite{Barnich:2010eb,Barnich:2011mi}, and applied to horizon phase space in~\cite{Adami:2021nnf,Odak:2023pga}.},
\begin{align}\label{Energy Smarr}
	E
	=
	2\left(
	\frac{\kappa_\kerr{\scriptstyle{(S_\dyn,J)}}}{2\pi}S_\dyn
	+\Omega_\kerr{\scriptstyle{(S_\dyn,J)}}J
	\right).
\end{align}
This definition agrees with the Kerr mass, and hence with the ADM
energy, on every stationary Kerr configuration. Moreover, for an
arbitrary variation in the dynamical horizon phase space, the chain
rule gives the generalized first law
\begin{align}\label{first law}
	\de E
	=
	\frac{\kappa_\kerr{\scriptstyle{(S_\dyn,J)}}}{2\pi}\de S_\dyn
	+\Omega_\kerr{\scriptstyle{(S_\dyn,J)}}\de J\,.
\end{align}
Note that $\kappa_\kerr$ and $\Omega_\kerr$ are the thermodynamic potentials determined
by the Kerr equation of state; away from the Kerr family they need not
coincide with the geometrical inaffinity and angular velocity of the
dynamical horizon.

The energy flux follows from \eqref{first law},
\begin{align}\label{energy flux}
\frac{\dd E}{\dd\hv}
	=
	\frac{\kappa_\kerr{\scriptstyle{(S_\dyn,J)}}}{2\pi} \frac{\dd S_\dyn}{\dd\hv}
	+\Omega_\kerr{\scriptstyle{(S_\dyn,J)}}\,\frac{\dd J}{\dd\hv}\,.
\end{align}
which can be expanded using \eqref{Sdyn flux} and \eqref{Y flux}.

\subsection{Perturbative expansion of charges and fluxes}
Expand the induced metric as 
\begin{align}
	q_{AB}=\bq_{AB}+\eps q^\bo{1}_{AB}+\eps^2 q^\bo{2}_{AB}+ \cO(\eps^3),
\end{align}
Let	$q_\po{n}\equiv	\bar q^{AB}q^\bo{n}_{AB}$, The area density is expanded as 
\begin{align}
	\sqrt q =\sqrt{\bq}	\left[	1+\eps \mu_\po{1}+\epsilon^2\mu_\po{2}	\right]	+ \cO(\eps^3),
\end{align}
where $\mu_\po{1}=\frac12 q_\po{1}$ and $\mu_\po{2}	\equiv	\frac12q_\po{2}	+\frac18q_\po{1}^2	-\frac14q^\bo{1}_{AB}q_\po{1}^{AB}$. The shape tensor density $\gamma_{AB}$ is expanded as
\begin{align}
	\gamma_{AB}=\frac{1}{\sqrt{\bq}}\left[\bq_{AB}+\eps\left(q^\bo{1}_{AB}-\mu_\po{1} \bq_{AB}\right)+\eps^2\left(q^\bo{2}_{AB}-\mu_\po{1} q^\bo{1}_{AB}+\bq_{AB}(\mu_\po{1}^2-\mu_\po{2})\right)\right]+ \cO(\eps^3),
\end{align}
Through second order, the deformation tensor is expanded as $\Theta_{AB}	=\eps \Theta^\bo{1}_{AB}+\eps^2 \Theta^\bo{2}_{AB}$ with 
\begin{align}
	\Theta^\bo{1}_{AB}	&=	\frac12\partial_{\hat v}q^\bo{1}_{AB},\qquad	\Theta^\bo{2}_{AB}=	\frac12	\left(\partial_{\hat v}q^\bo{2}_{AB}+\partial_{\tilde v}q^\bo{1}_{AB}\right),
\end{align}
and hence $\th=\eps^2 \th_2$, and $\sigma_{AB}=\eps \sigma^\bo{1}_{AB}+\eps^2 \sigma^\bo{2}_{AB}$ with 
\begin{subequations}
\begin{align}
		\theta_\po{2}	&=	\bar q^{AB}\Theta^\bo{2}_{AB}-q_\po{1}^{AB}\Theta^\bo{1}_{AB},\\
	\sigma^\bo{1}_{AB}&=	\Theta^\bo{1}_{AB},	\\
	\sigma^\bo{2}_{AB}	&=	\Theta^\bo{2}_{AB}	-\frac12\theta_\po{2}\bar q_{AB},\\
	\sigma^{AB}	&=	\epsilon\sigma_\po{1}^{AB}	+\epsilon^2\Sigma_\po{2}^{AB}\,,\qquad \Sigma_\po{2}^{AB}\equiv\sigma_\po{2}^{AB}-2q_\po{1}^{C(A}{\sigma^\bo{1}}^{B)}{}_{C}
\end{align}
\end{subequations}
\subsubsection{Charges to second order}

The charges expanded to second order read explicitly 
\begin{subequations}
	\begin{align}
		Q_{ T}	&=	\frac12	\oint\sqrt{\bar q}\, T	\left[	\bar\kappa	+\frac{\epsilon}{2}\bar\kappa q_\po{1}	+\epsilon^2	\left(	\bar\kappa\mu_\po{2}	-\theta_\po{2}\right)\right]\\
		Q_{ Y}&	=	\frac12	\oint\sqrt{\bar q}\,Y^A	\left[	\bar\omega_A	+\epsilon		\left(		\omega_A^\bo{1}	+\frac12q_\po{1}\bar\omega_A	\right)	+\epsilon^2	\left(	\omega_A^\bo{2}+\frac12q_\po{1}\omega_A^\bo{1}+\mu_\po{2}\bar\omega_A			\right)\right]\\
		Q_{W}&=-\frac{\eps^2}{2}	\oint\sqrt{\bar q}\,W e^{-\bk\hv}\th_\po{2}\,.
\end{align}
\end{subequations}
For $ T=1$, this becomes
\begin{align}
		\begin{aligned}
			Q_{ T=1}
			&=
			\frac{1}{8\pi}
			\Big[
			\bar\kappa\bar A
			+\epsilon\bar\kappa A_\po{1}
			+\epsilon^2
			\left(
			\bar\kappa A_\po{2}
			-\partial_{\hat v}A_\po{2}
			-\partial_{\tilde v}A_\po{1}
			\right)
			\Big]
			+\mathcal O(\epsilon^3).
		\end{aligned}
\end{align}
where $A=\bar{A}+\eps A_\po{1}+\eps^2 A_\po{2}+\cO(\eps^3)$ is the perturbative expansion of the area of the horizon. Note that $\ddhv A=\int d^2x \sqrt{q}\,\th$ with \eqref{th leading} implies
\begin{align}
	\pd_\hv A_\po{1}=0\,,\qquad \pd_\tlv A_\po{1}+\pd_\hv A_\po{2}=\int \sqrt{\bq}\,\th_\po{2}
\end{align}
which are solved by 
\begin{align}
	A=\bar{A}+\eps A_\po{1}(\cv)+\cO(\eps^2)\,,\qquad \pd_\cv A_\po{1}(\cv)=\int_{S_\tlv}\sqrt{\bq}\avg{\th_\po{2}}\,.
\end{align}
where $\cv=\eps \hv$ is the adiabatic time of the Carter coordinates. 
\subsubsection{Fluxes to third order}
To third order, the fluxes read
\begin{align}
			\cF_{f}&=\frac{\epsilon^2}{2}
			\oint\sqrt{\bar q}\, f\,\left[
			\sigma^\bo{1}_{AB}\,\sigma_\po{1}^{AB}+\eps
			\Big(2\sigma_\po{1}^{AB}\,\sigma^\bo{2}_{AB}-2q_\po{1}^{AC}	\sigma^\bo{1}_{AB}\,{\sigma_\po{1}}^B{}_{C}	+\frac12q_\po{1}	\sigma^\bo{1}_{AB}\,\sigma_\po{1}^{AB}\Big)	\right]
\end{align}
For Lorentz transformations generated by \eqref{Lorentz generator}, we can extend the flux formula \eqref{Lorenz flux 2ndO} to third order
\begin{align}\label{Lorenz flux 3rdO}
	\cF_{\chi}&=\frac{\eps^2}{4}\oint \bar{\mu} \left[\bar{\mu}\sigma_\po{1}^{AB}\cL_\bY \gamma^\bo{1}_{AB}-\bsi\, \pd_\hv \th_\po{2}/\bk\right]\nonumber\\
	&+\frac{\eps^3}{4}\oint \bar{\mu}\bigg[\bar{\mu} \Big(\sigma_\po{1}^{AB}\cL_\bY \gamma^\bo{2}_{AB}+\Sigma_\po{2}^{AB}\cL_\bY \gamma^\bo{1}_{AB}+2\mu_\po{1}\sigma_\po{1}^{AB}\cL_\bY \gamma^\bo{1}_{AB}\Big)\nonumber\\
	&\hspace{2cm} -\frac{\bsi}{\bk}\left(\pd_\hv\th_\po{3}+\pd_\tlv \th_\po{2}\right)+\mu_\po{1}\sigma_{(1)}^2\frac{\bsi}{\bk}-\th_\po{2}(\cL_\bY+\bsi)\mu_\po{1}\bigg]\,.
\end{align}

\paragraph{Angular momentum flux.} For a background rotational Killing vector $\phi$, the corresponding divergence $\bsi=\bD_A \phi^A=0$ and thus
\begin{align}\label{angular momentum flux 3rdO}
	\cF_{\chi}&=\frac{\eps^2}{4}\oint \bar{\mu} \left[\bar{\mu}\sigma_\po{1}^{AB}\cL_\bY \gamma^\bo{1}_{AB}\right]\nonumber\\
	&+\frac{\eps^3}{4}\oint \bar{\mu}\left[\bar{\mu} \left(\sigma_\po{1}^{AB}\cL_\bY \gamma^\bo{2}_{AB}+\Sigma_\po{2}^{AB}\cL_\bY \gamma^\bo{1}_{AB}+2\mu_\po{1}\sigma_\po{1}^{AB}\cL_\bY \gamma^\bo{1}_{AB}\right)- \th_\po{2}\,\cL_\bY\mu_\po{1}\right]\,.
\end{align}

\paragraph{Order-mixing invariance of the global charges.}
We can now assess the order-mixing ambiguity discussed in
Sec.~\ref{sec:finite corotating transformation}. Since the relative
transformation \eqref{order mixing residual} starts at $\cO(\eps^2)$,
all first-order horizon data are unchanged. From
\eqref{order mixing data}, we find
\begin{align}\label{mu2 ambiguity}
	\Delta_\zeta\mu_\po{2}
	=
	\frac12\bq^{AB}\Delta_\zeta q^\bo{2}_{AB}
	=
	\bD_A Z^A,
	\qquad
	\Delta_\zeta\theta_\po{2}=0,
\end{align}
where $\Delta_\zeta$ denotes the ambiguity in the Carter coordinates due to our ignorance of the 3rd order perturbation of the metric.
It follows that the ambiguity in the second-order supertranslation
charge is
\begin{align}\label{QT order mixing}
	\Delta_\zeta Q_T^{(2)}
	&=
	\frac{\bk}{2}
	\oint\sqrt{\bq}\,
	T\,\bD_A Z^A=
	-\frac{\bk}{2}
	\oint\sqrt{\bq}\,
	Z^A\bD_A T .
\end{align}
Thus the second order charge corresponding to a generic angle-dependent parameter $T(\hx^A)$ is ambiguous due to order mixing. The constant mode,
however, is protected:
\begin{align}\label{QT1 order mixing}
	\Delta_\zeta Q_{T=1}^{(2)}=\frac{2\pi}{\bk}\Delta_\zeta S_\dyn=0 .
\end{align}
Consequently, the dynamical-entropy charge is unambiguously determined through second order by a second-order bulk
perturbation.

The angular charges behave similarly. Using
\eqref{order mixing data}, their second-order ambiguity is
\begin{align}\label{QY order mixing}
	\Delta_\zeta Q_Y^{(2)}
	=
	\frac12\oint\sqrt{\bq}\,Y^A
	\left[
	\cL_Z\bar\omega_A
	+\bk\bD_A F
	+\big(\bD_BZ^B\big)\bar\omega_A
	\right].
\end{align}
Integrating by parts, this may be written as
\begin{align}\label{QY order mixing 2}
	\Delta_\zeta Q_Y^{(2)}
	=
	\frac12\oint\sqrt{\bq}\,
	[Y,Z]^A\bar\omega_A
	-\frac{\bk}{2}
	\oint\sqrt{\bq}\,
	F\,\bD_A Y^A .
\end{align}
Hence a generic superrotation charge, and in particular a generic Lorentz
charge, is not protected against order mixing.
However, the angular momentum associated to the background axial Killing symmetry is an important exception. For
$Y=\phi$, axisymmetry of the background implies
\begin{align}
	\bD_A\phi^A=0,
	\qquad
	\cL_\phi\bq_{AB}=0,
	\qquad
	\cL_\phi\bar\omega_A=0 .
\end{align}
The second term in \eqref{QY order mixing 2} therefore vanishes. Moreover,
on the closed horizon cut,
\begin{align}\label{J invariance}
	\Delta_\zeta J^{(2)}&=-\frac12 \oint\sqrt{\bq}\,
	[\phi,Z]^A\bar\omega_A
	=-\frac12 
	\oint \cL_\phi
	\left(\sqrt{\bq}\,Z^A\bar\omega_A\right)=0\,.
\end{align}
The axial angular momentum is therefore also unambiguously determined through second order.
Finally, the horizon energy obeys \eqref{first law}, and therefore 
\begin{align}\label{EH order mixing}
	\Delta_\zeta E^{(2)}=0 \,.
\end{align}
upon using \eqref{QT1 order mixing},\eqref{J invariance}.
Thus, although the local Carter data $q^\bo{2}_{AB}$ and
$\omega_A^\bo{2}$ depend on the third-order completion of the
coordinate transformation, the three global Killing charges of primary
interest here---the dynamical entropy, axial angular momentum, and
horizon energy---do not. Their second-order values can be
obtained unambiguously from a second-order bulk perturbation.

\section{Summary}
The paper is summarized in Figure \ref{fig:summary} below. The flowchart can be used as a guideline to implement horizon fluxes in the modeling of EMRI waveforms. In the future, we plan to apply this construction on available EMRI metric perturbation to extract the evolution of horizon parameters. We also note that the construction of charges and fluxes relied on Carter coordinates on the horizon. It would be useful to formulate horizon flux-balance laws in a generic horizon coordinate system. 
\begin{figure}[H]
	\centering
	\begin{tikzpicture}[
		font=\footnotesize,
		node distance=5mm and 8mm,
		>={Latex[length=2mm]},
		arrow/.style={
			->,
			line width=.65pt
		},
		banner/.style={
			draw,
			rounded corners=2pt,
			line width=.7pt,
			fill=black!4,
			text width=.88\linewidth,
			align=center,
			inner xsep=3pt,
			inner ysep=5pt
		},
		header/.style={
			draw,
			rounded corners=2pt,
			fill=black!12,
			text width=.40\linewidth,
			align=center,
			inner xsep=2pt,
			inner ysep=3pt,
			font=\small\bfseries
		},
		stage/.style={
			draw,
			rounded corners=2pt,
			fill=black!3,
			text width=.40\linewidth,
			align=left,
			minimum height=2.15cm,
			inner sep=6pt
		},
		output/.style={
			draw,
			rounded corners=2pt,
			fill=black!8,
			text width=.88\linewidth,
			align=left,
			inner sep=7pt
		}
		]
		
		\node[banner, inner xsep=1pt] (banner) {%
			\textbf{Automatable framework to include horizon fluxes for EMRI/LISA waveform modeling:}\\[1mm]
			 Bulk perturbative solution
			$\rightarrow$ Gauge-transformation 
			$\rightarrow$ Extraction of horizon geometry
			$\rightarrow$ Computation of horizon fluxes 
			$\rightarrow$ Determine slow evolution of the primary  
			$\rightarrow$ waveform correction
		};
		
		\node[header, anchor=north east] (geom-h)
		at ([xshift=-4mm,yshift=-6mm]banner.south) {%
			Horizon dynamics
		};
		
		\node[header, anchor=north west] (pipe-h)
		at ([xshift=4mm,yshift=-6mm]banner.south) {%
			Bulk-to-horizon pipeline
		};
		
		\node[stage, below=of geom-h] (inu) {%
			\textbf{1. Horizon-adapted coordinates.}
			Construct the ingoing Newman--Unti gauge and its near-horizon expansion:
			Eqs.~\eqref{INU gauge}--\eqref{Bondi variables}.
			The fully adapted Carter form is Eq.~\eqref{NH Carter}.
		};
		
		\node[stage, below=of inu] (dynamics) {%
			\textbf{2. Horizon geometry and exact dynamics.}
			The data are
			$\{q_{AB},\kappa,V^A,\omega_A,\Theta_{AB}\}$;
			see Eqs.~\eqref{kinematic variables} and \eqref{NH data}.
			Their exact propagation is governed by the
			Raychaudhuri--Damour--shear tower~\eqref{evolution tower}.
		};
		
		\node[stage, below=of dynamics] (multiscale) {%
			\textbf{3. Two-timescale reduction.}
			Introduce $\tilde v=\epsilon v$ and
			$\partial_v\mapsto
			\partial_v+\epsilon\partial_{\tilde v}$,
			Eq.~\eqref{v der}, together with the flow average
			\eqref{averaging def} and its properties
			\eqref{averaging properties}.
		};
		
		\node[stage, below=of multiscale] (rigidity) {%
			\textbf{4. Coarse-grained constraints.}
			Multiscale analysis of horizon equations \eqref{evolution tower} reveals \textit{adiabatic rigidity}
			\[
			\kappa_1=\kappa_1(\tlv),\;\;
			\langle\sigma^{(1)}_{AB}\rangle=0,\;\;
			V_1^A=\Omega_1(\tilde v)\phi^A .
			\]
			See Eqs.~\eqref{th leading}, \eqref{zero avg shear},
			\eqref{V1 Kerr}, \eqref{kappa1 multiscale}.
		};
		
		\node[stage, below=of pipe-h] (input) {%
			\textbf{1. Input: user-gauge perturbation.}
			Start from a bulk perturbation in a generic user gauge,
			Eq.~\eqref{metric perturbed}, with
			$\bar g^{\mu\nu}$ assumed to be already expressed in INU coordinates.
		};
		
		\node[stage, below=of input] (gauge) {%
			\textbf{2. Transform to INU gauge.}
			Apply the knight transformation
			\eqref{knight} and \eqref{metric gauge transf}.
			Solve the bulk and boundary gauge-fixing equations
			\eqref{NU gauge fixing} and \eqref{horizon gauge eqs}.
		};
		
		\node[stage, below=of gauge] (extract) {%
			\textbf{3. Read off the horizon data.}
			Extract $V^A,q_{AB},\kappa,\omega_A$ directly from the INU metric
			using \eqref{linear data INU}.
			Explicit first-order expressions are given in
			\eqref{induced metric 1st order}--\eqref{Hajicek 1st order};
			second order follows from \eqref{2nd order data} and
			\eqref{2nd order deformation}.
		};
		
		\node[stage, below=of extract] (carter) {%
			\textbf{4. Transform to Carter coordinates.}
			Use the INU-to-Carter map \eqref{coord transf},
			transform the null frame according to \eqref{tetrad transform},
			and obtain the Carter data from \eqref{transformed data}.
			The finite transformation exhibits order mixing in
			$\hat q_{AB}$ and $\hat\omega_A$. Global Killing charges are unaffected.
		};
		
		\draw[arrow] (inu) -- (dynamics);
		\draw[arrow] (dynamics) -- (multiscale);
		\draw[arrow] (multiscale) -- (rigidity);
		
		\draw[arrow] (input) -- (gauge);
		\draw[arrow] (gauge) -- (extract);
		\draw[arrow] (extract) -- (carter);
		
		\coordinate (merge) at
		($(rigidity.south)!0.5!(carter.south)+(0,-8mm)$);
		
		\node[output, anchor=north] (charges) at (merge) {%
			\textbf{5. Residual symmetries, charges, and fluxes.}
			Carter-gauge symmetries are given by Eq.~\eqref{symmetries},
			and the associated charges by \eqref{charges def}.
			Their exact balance law is Eq.~\eqref{flux v1}
			or equivalently the shape--area form
			\eqref{flux shape area}.
			The principal global quantities are the dynamical entropy
			\eqref{Q T=1}--\eqref{Sdyn flux}, Lorentz charges
			\eqref{Lorentz generator}, angular momentum \eqref{J def},
			and the energy obeying the first law \eqref{first law}.
			Charges are expanded through $\mathcal O(\epsilon^2)$ and
			fluxes through $\mathcal O(\epsilon^3)$; see
			Eqs.~\eqref{Lorenz flux 3rdO} and
			\eqref{angular momentum flux 3rdO}.
		};
		
		\path let
		\p1=(rigidity.south),
		\p2=(charges.north)
		in
		coordinate (rigidity-to-charges) at (\x1,\y2);
		
		\path let
		\p1=(carter.south),
		\p2=(charges.north)
		in
		coordinate (carter-to-charges) at (\x1,\y2);
		
		\draw[arrow]
		(rigidity.south) -- (rigidity-to-charges);
		
		\draw[arrow]
		(carter.south) -- (carter-to-charges);
		
	\end{tikzpicture}
	
	\caption{This figure serves as the summary of the paper. The left branch summarizes the
		intrinsic geometric and multiscale analysis of the horizon, while the
		right branch gives the constructive map from a bulk perturbation in a user
		gauge to the horizon data in Carter coordinates. The two branches meet in
		the computation of charges and flux-balance laws. The pipeline is suitable for symbolic or numerical automation and may be used for high-accuracy EMRI waveform models for LISA.}
	\label{fig:summary}
\end{figure}
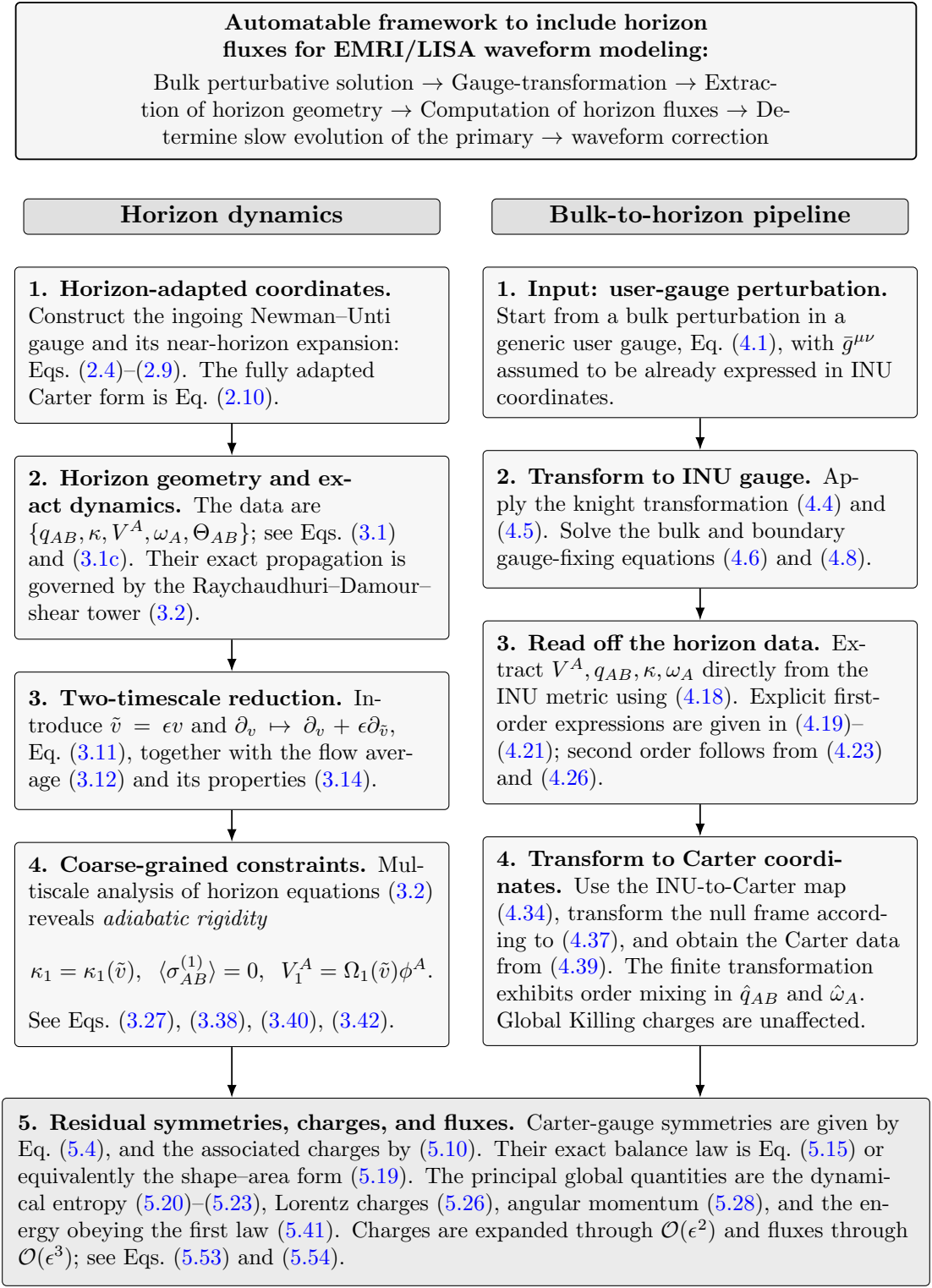

\paragraph{Acknowledgments.} This work would not have been possible without the contribution of Adam Pound. I appreciate his constant support. I am also grateful to L. Ciambelli, A. Grant, A. Ribes Metidieri, S. Speziale, A. Spiers and L. Stein for useful related discussions.

\appendix

\section{Green function on the horizon}\label{app: Green}
On the event horizon, equations of the form 
\begin{align}\label{ODE}
	(\pd_v-\bk)f=-g
\end{align}
are ubiquitous. Its exact solution, can be found by a suitable Green function, solving $(\pd_v-\bk)G(v-v')=-\de(v-v')$. The retarded and advanced Green functions are
\begin{align}
	G_{\rm R}(v-v')=-e^{\bk(v-v')}\Th(v-v')\,,\qquad G_{\rm A}(v-v')=e^{\bk(v-v')}\Th(v'-v)
\end{align}
where $\Th(v)$ is the step function. The corresponding solutions are
\begin{align}\label{R-A sols}
	f_{\rm R}(v)=-\int_{-\infty}^v e^{\bk(v-v')}g(v')dv'\,,\qquad f_{\rm A}(v)=\int_v^\infty e^{-\bk(v'-v)}g(v')dv'.
\end{align}
The retarded solution exponentially diverges at late time, even if the source is turned on for a finite interval. On the other hand, the advanced solution is well behaved at late time for reasonable source $g$. We are therefore led to select the advanced solution. The price to pay is that this solution is teleological, \ie its state at $v$ is dependent on its future evolution. This is conceivable, as the event horizon is a global concept, separating null rays that can and those that cannot reach infinity. We refer the reader to~§6 of \cite{Thorne:1986iy} for further discussion.

\paragraph{Time localization.} The teleological nature of the above solution is not as disturbing as it may sound. The exponential function in the advanced solution in \eqref{R-A sols} suppresses dependence on events far in the future. As a result, the solution can be localized in time in a perturbative treatment. To see this, write the solution to \eqref{ODE} formally as 
\begin{align}
	f=\frac{1}{\bk}\big[1-\pd_v/\bk\big]^{-1} g. 
\end{align}
Note that $\frac{\pd_v}{\bk} g\sim \frac{r_\circ}{\lambda}g$, where $r_\circ$ is the radius of the black hole, and $\lambda$ is the typical wavelength of the source function $g$. As long as $\frac{1}{g}\frac{\pd_v g}{\bk}\sim \frac{r_\circ}{\lambda}\ll1$, we can Taylor expand $\big[1-\pd_v/\bk\big]^{-1}$ to arrive at~\cite{Poisson:2004cw,Bonetto:2021exn}
\begin{align}
	f(v)=\frac{g(v)}{\bk}+\frac{\pd_v g(v)}{\bk^2}+\cdots.
\end{align}
Thus $f(v)$ can be approximated through $g$ and its few derivatives at the same time $v$. 

\bibliography{Refs}

\end{document}